\documentclass[twocolumn, times]{aastex631}

\usepackage{color}

\defcitealias{Bellini2014}{Paper~1}
\defcitealias{Watkins2015a}{Paper~2}
\defcitealias{Watkins2015b}{Paper~3}
\defcitealias{Baldwin2016}{Paper~4}
\defcitealias{Libralato2018}{Paper~6}

\newcommand {\Msun} {M$_\odot$}

\newcommand {\Nused} {N_\mathrm{used}}
\newcommand {\Nfound} {N_\mathrm{found}}

\newcommand {\Ri} {R_\mathrm{i}}
\newcommand {\mi} {m_\mathrm{i}}

\newcommand {\meq} {m_\mathrm{eq}}
\newcommand {\logmeq} {\log_{10} \meq}
\newcommand {\sigmaeq} {\sigma_\mathrm{eq}}
\newcommand {\ms} {m_\mathrm{s}}
\newcommand {\sigmas} {\sigma_\mathrm{s}}
\newcommand {\ck} {c_k}

\newcommand {\Ncore} {N_\mathrm{core}}

\newcommand {\Nhalf} {N_\mathrm{half}}

\newcommand {\Rhalf} {R_\mathrm{half}}

\newcommand {\FeH} {[Fe/H]}

\graphicspath{{./}{figures/}}

\shorttitle{HSTPROMO GCs. VII. Energy Equipartition}
\shortauthors{Watkins et al.}

\begin{document}

\title{\textit{Hubble Space Telescope} Proper Motion (HSTPROMO) Catalogs of Galactic Globular Clusters. VII. Energy Equipartition}

\correspondingauthor{Laura Watkins}
\email{lwatkins@stsci.edu}

\author[0000-0002-1343-134X]{Laura L. Watkins}
\affiliation{AURA for the European Space Agency, ESA Office, Space Telescope Science Institute, 3700 San Martin Drive, Baltimore MD 21218, USA}
\affiliation{European Southern Observatory, Karl-Schwarzschild-Stra{\ss}e 2, D-85748 Garching bei M{\"u}nchen, Germany}
\affiliation{Department of Astrophysics, University of Vienna, T{\"u}rkenschanzstra{\ss}e 17, 1180 Vienna, Austria}

\author[0000-0001-7827-7825]{Roeland P. van der Marel}
\affiliation{Space Telescope Science Institute, 3700 San Martin Drive, Baltimore MD 21218, USA}
\affiliation{Center for Astrophysical Sciences, Department of Physics \& Astronomy, Johns Hopkins University, Baltimore MD 21218, USA}

\author[0000-0001-9673-7397]{Mattia Libralato}
\affiliation{AURA for the European Space Agency, ESA Office, Space Telescope Science Institute, 3700 San Martin Drive, Baltimore MD 21218, USA}

\author[0000-0003-3858-637X]{Andrea Bellini}
\affiliation{Space Telescope Science Institute, 3700 San Martin Drive, Baltimore MD 21218, USA}

\author[0000-0003-2861-3995]{Jay Anderson}
\affiliation{Space Telescope Science Institute, 3700 San Martin Drive, Baltimore MD 21218, USA}

\author[0000-0002-1212-2844]{Mayte Alfaro-Cuello}
\affiliation{Space Telescope Science Institute, 3700 San Martin Drive, Baltimore MD 21218, USA}

\begin{abstract}

We examine the degree of energy equipartition in 9 Galactic globular clusters using proper motions measured with the \textit{Hubble Space Telescope}. For most clusters in the sample, this is the first energy equipartition study ever performed. This study is also the largest of its kind, albeit with only 9 clusters. We begin by rigorously cleaning the catalogues to remove poor-quality measurements and to ensure high signal-to-noise for the study. Using the cleaned catalogues, we investigate how velocity dispersion $\sigma$ changes with stellar mass $m$. We fit two functional forms: the first, a classic power-law of the form $\sigma \propto m^{-\eta}$ where $\eta$ is the degree of energy equipartition, and the second from \citet{Bianchini2016} parameterised by an equipartition mass $\meq$ where $\eta$ changes with stellar mass. We find that both functions fit well but cannot distinguish with statistical significance which function provides the best fit. All clusters exhibit varying degrees of partial equipartition; no cluster is at or near full equipartition. We search for correlations of $\eta$ and $\meq$ with various cluster properties. The most significant correlation is observed with the number of core or median relaxation times ($\Ncore$ or $\Nhalf$) the cluster has experienced. Finally, we determine the radial equipartition profile for each cluster, that is, how the degree of equipartition changes with projected distance from the cluster centre. We do not detect statistically significant trends in the degree of equipartition with radius. Overall, our observational findings are in broad agreement with theoretical predictions from \textit{N}-body models published in recent years.

\end{abstract}


\section{Introduction}
\label{section:introduction}

Globular clusters (GCs) are collisional systems, meaning that the stars inside them interact on timescales significantly shorter than the age of the Universe. This, combined with the fact that they are very old (a significant fraction of the age of the Universe), means that the stars have undergone a lot of interactions. During an interaction, stars will mildly perturb each other's orbits and exchange a small amount of energy, in a process known as two-body relaxation. Although these effects are small individually, they compound and the consequences can be considerable after many interactions \citep{Spitzer1987}.

One such consequence is energy equipartition. During an interaction, energy will generally be transferred from the star with higher energy to the star with lower energy \citep{Spitzer1969}. So given enough time (or interactions), and if no other factors are at work, we would expect that an ensemble of stars would reach a state where they all have the same kinetic energy $\frac{1}{2} m \sigma^2$, for stellar mass $m$ and velocity dispersion $\sigma$. Or, alternatively, $\sigma \propto m^{-0.5}$. Thus, we would observe that low-mass stars move faster than high-mass stars.

This describes a hypothetical end state. In practice, it's likely that a stellar system may be in partial equipartition where there is some dependence on stellar mass in the velocity distribution, but it has not reached full equipartition. Classically, we consider that dispersion $\sigma$ changes with stellar mass $m$ like
\begin{equation}
    \sigma = \sigma_s \left( \frac{m}{m_s} \right)^{-\eta}
    \label{equation:equipartition}
\end{equation}
where $m_s$ is a scale mass, $\sigma_s$ is the velocity dispersion of a star of mass $m_s$ and $\eta$ quantifies the degree of energy equipartition. $\eta = 0$ indicates no equipartition, that is the velocities are independent of stellar mass. $\eta = 0.5$ indicates full energy equipartition whereby all stars have the same kinetic energy.

Whether equipartition is partial or full, we would expect slow-moving high-mass stars to be more affected by dynamical friction than fast-moving low-mass stars. As a result, the high-mass stars would tend to sink towards the centre of a cluster and low-mass stars would tend to move outwards. This is known as mass segregation. Studies are challenging as they require accurate star counts for both bright and faint stars, but are possible from space and from the ground \citep[e.g.][]{King1995, Anderson1997, Koch2004, Heyl2012, Dalessandro2015, Sollima2017}. Recently, \citet{Baumgardt2022} have even suggested that mass segregation could be a discriminator between GCs and ultra-faint dwarfs.

Mass segregation means that stars of different mass have different spatial distributions, and we know also the velocity dispersion changes as a function of position within the cluster. So complete equipartition will only be indicated by $\eta = 0.5$ over small radial ranges, and may not hold globally.

Energy equipartition is even more challenging to study. The reason for this is that we need to measure high-precision kinematics for stars that span a wide range of stellar masses. Significant mass loss can occur on the RGB, but these end stages of stellar evolution are so fast compared to the 2-body interaction timescales, that the stars do not have time to adjust kinematically to their new mass. So even though there is an intrinsic mass range from the main-sequence turn-off (MSTO) up along the subgiant branch and red giant branch, those stars behave like a population spanning a relatively small range of stellar masses. Where we see the wide ranges of stellar masses that we need for this kind of work is along the (fainter) main sequence. Unfortunately, the stars for which we have historically been able to measure kinematics (at all, let alone with the precision required) are the brightest stars. This is true for both spectroscopic studies that have measured line-of-sight velocities (LOSVs) and astrometry studies that have measured proper motions (PMs).

There are a few exceptions. Blue stragglers are stars both bluer and brighter than the MSTO that sit on the high-mass main sequence. They are thought to be once-lower-mass main sequence stars that have recently gained extra mass via collision or mass-transfer from an evolved companion in a binary system. Their original mass explains their longevity and their new mass explains their location in the HR-diagram. These are a rare but useful population of stars that are both bright and of a different mass than the rest of the bright stars in a cluster; they can even be used as dynamical `clocks' \citep[see][for a nice review]{Ferraro2020}. \citet{Baldwin2016} studied the kinematic differences between blue straggler stars and the evolved stars in 19 clusters using \textit{HST} PMs and found that the blue straggler stars are indeed moving slower on average, indicating that there is some degree of energy equipartition and that they are of higher mass.

The observational challenges have not prevented theoretical work on this topic. \citet{Baumgardt2003} studied internal GC dynamics using \textit{N}-body simulations and found that GC centres can reach full equipartition after core collapse but that the outer regions never reach full equipartition unless it is present initially.

\citet{Trenti2013} studied energy equipartition using \textit{N}-body simulations of GCs. To match observations, they studied equipartition using projected velocity dispersions and found that their simulated clusters never reached full energy equipartition in any region. The simulated clusters started with no equipartition ($\eta = 0$) and were allowed to evolve. The inner regions developed some velocity dependence on stellar mass quite rapidly, reaching a maximum of $\eta \sim 0.2$ and then actually turned over and $\eta$ decreased to around $\eta \sim 0.1$. The outer regions steadily increased from $\eta = 0$ to $\eta = 0.1$ without the turn over. This is remarkably different behaviour than our somewhat naive expectation that equipartition would increase until complete. Another prediction of this work is that equipartition (or $\eta$) would be highest in the central regions of the cluster and then would decrease with increasing distance from the cluster centre. These are two quantifiable results that we would like to put to the test: what values of $\eta$ do we observe in real clusters? And how does $\eta$ change with radius?

\citet{Bianchini2016} also studied equipartition in \textit{N}-body simulations of GCs, also in projection. They found that not only does velocity dispersion depend on stellar mass, so does equipartition itself. They offered a new parameterisation for velocity dispersion as a function of stellar mass,
\begin{equation}
	\sigma(m) = \left\{ \begin{array}{cc}
    	\sigma_0 \exp \left( - \frac{1}{2} \frac{m}{\meq} \right) & \qquad m \le \meq \\
        \sigmaeq \left( \frac{m}{\meq} \right)^{-\frac{1}{2}} & \qquad m > \meq
    \end{array} \right. ,
    \label{equation:Bianchini}
\end{equation}
where $\meq$ is a mass scale parameter that quantifies the level of energy equipartition in a system. $\sigma_0 = \sigma(0)$ and $\sigmaeq = \sigma(\meq)$ are scale parameters that are related such that $\sigmaeq = \sigma_0 \exp \left( - \frac{1}{2} \right)$. In the classic power-law function, $\eta$ is the derivative of the dispersion with stellar mass. So it is useful to consider the derivative of this function as well,
\begin{equation}
	\eta(m) = - \frac{d \ln \sigma}{d \ln m} = \left\{ \begin{array}{cc}
    	\frac{1}{2} \frac{m}{\meq} & \qquad m \le \meq \\
        \frac{1}{2} & \qquad m > \meq
    \end{array} \right.
    \label{equation:BianchiniEta}
\end{equation}
This helps us interpret $\meq$. Under this formalism, stars more massive than $\meq$ are in complete equipartition, and stars less massive are in partial equipartition, with the degree of equipartition decreasing with stellar mass. This offers another quantifiable test for us to investigate: how well does this functional form describe real clusters? Is it preferred over the classic power-law description?

\citet{Anderson2010} measured the degree of equipartition in the inner regions of NGC\,5139 ($\omega$ Centauri) using a catalogue of PMs measured over a 4-year baseline with \textit{HST}. They estimated $\eta \sim 0.2$, in good agreement with the later theoretical study of \citet{Trenti2013}. The latter also quote a revised estimate (via private communication) for the equipartition in NGC\,5139 using an improved \textit{HST} PM catalogue of $\eta \sim 0.16$, again in good agreement with their theoretical predictions. That improved PM catalogue for NGC\,5139, along with PM catalogues for 21 other GCs (including NGC\,362), was later published in \citet{Bellini2014}. It is these catalogues that form the basis of the present work.

\citet{Libralato2018} presented a further-improved catalogue for NGC\,362, with which they studied energy equipartition in the cluster. They found an average degree of equipartition $\eta = 0.114 \pm 0.012$, again consistent with theoretical expectations. They also found that $\eta$ is highest near the cluster centre ($\eta \sim 0.4$) and decreases with increasing radius (to $\eta \sim 0.1$), again consistent with \citet{Trenti2013}. Improved catalogues are forthcoming for a number of other clusters, but at present, the largest catalogue sample remains that presented in \citet{Bellini2014}\footnote{We use the original, not improved, catalogue for NGC\,362; we wish to use catalogues that were generated in the same way so that we can analyse them all consistently.}.

In our initial studies of these catalogues -- which focused on velocity dispersion and anisotropy profiles \citep{Watkins2015a}, and dynamical distances and mass-to-light ratios \citep{Watkins2015b} -- we made a magnitude cut at the MSTO and restricted our analysis to only the bright stars. The reasons were twofold: first, to mitigate the same dependence of kinematics on stellar mass that we are seeking in this work, and second, to focus on the stars that were generally easier to measure. Assessing the quality of just the bright-star catalogues was already challenging. Moreover, most clusters had never been studied with PMs before and never had such a large sample of GCs been studied with PMs as an ensemble, so even analysing the bright stars represented a big step forwards in our understanding.

However, it is now time to restudy these catalogues in their entirety, with no magnitude cut. By including the main-sequence stars, we significantly increase the stellar mass range spanned by these catalogues, and make it feasible to study equipartition and attempt to answer some of the questions posed by earlier theoretical work. Although this has been done for a handful of GCs individually, this is the first time such a study has been done for an ensemble.

As for our earlier analysis \citep{Watkins2015a}, careful cleaning of the catalogues is crucial to ensure that we have a sample of high-quality well-measured stars in each cluster. This is more challenging here with the sample now spanning a broader magnitude range and with the complication of stellar mass in play. Although we start with 22 PM catalogues in total, only 9 clusters will prove to be of sufficient quality to complete the analysis.

In \autoref{section:masses}, we describe how we use the colour-magnitude information to estimate masses for all the stars in our sample. In \autoref{section:datacleaning}, we describe the modified cleaning process to select high-quality samples and identify clusters with sufficient quality to proceed. In \autoref{section:equipartition}, we investigate the energy equipartition in the 9 GCs that passed the quality selections.  In \autoref{section:etaprofiles}, we investigate how equipartition changes with distance from the cluster centre. In \autoref{section:discussion}, we search for correlations of equipartition parameters with other cluster properties and discuss the broader implications of our results. Finally, in \autoref{section:conclusions}, we summarise our conclusions.

\section{Stellar Masses}
\label{section:masses}

If we are to study the effect of stellar mass on cluster kinematics, we must first estimate masses for the stars in our catalogues. We do this via isochrone fitting, for which we use isochrones from the Dartmouth Stellar Evolution Database \citep{Dotter2008}.

Most clusters have F606W and F814W magnitudes, and to form the CMDs we use the F606W magnitude and the F606W-F814W colour. There are two exceptions. NGC\,5139 has F435W, F625W and F658N magnitudes, and to form the CMD we use the F625W magnitude and the F435W-F625W colour. NGC\,6266 has F390W and F658N magnitudes, and to the form the CMD, we use the F658N magnitude and the F390W-F658N colour.

We adopt cluster metallicities [Fe/H] and extinctions E(B-V) from \citet[][2010 edition]{Harris1996} and assumed primordial He. Where available, we use cluster ages from \citet{VandenBerg2013}; where not available, we infer the age by calculating the weighted average of the ages in the full \citet{VandenBerg2013} sample, using Gaussian weights in metallicity centred on the target cluster and with width 0.1~dex. Ideally, we would use alpha-element abundances of [$\alpha$/Fe] = 0.3 as is typical for GCs, but this value is not available in the isochrone set. So instead we adopt [$\alpha$/Fe]~=~0.2~dex for clusters with [Fe/H]~$>$~-1.5~dex and [$\alpha$/Fe]~=~0.4~dex for clusters with [Fe/H]~$\le$~-1.5~dex. Where possible, we use extinction coefficients from \citet{Casagrande2014}; for F625W and F658N, we infer the coefficients via interpolation using the coefficients in \citet{Casagrande2014} and central wavelengths for other ACS filters from \citet{Ryon2018}.

Before fitting isochrones, we wish to be sure that we have a high-quality sample of genuine cluster members for the baseline of the fits. The full data quality analysis is described in detail in \autoref{section:datacleaning}; here we perform a simpler version of the cuts, that is good enough for our purposes here as we only wish to clean up the CMD and select for membership, not rigourously select the members the best measured PMs. We first performed the steps outlined in \autoref{section:pmfitquality} and \autoref{section:positionfitquality}, and then crudely sigma-clip to remove obvious contaminants as described at the start of \autoref{section:localdispersions}. None of these steps depends on stellar mass.

We begin by selecting an isochrone with the given cluster properties and identify the point on the isochrone that best matches each star in the high-quality sample. To do this, we interpolate in magnitude to finely sample the isochrone within $\pm$0.5~dex of the target star in steps of 0.01~dex. We then calculate the likelihood of the star for each point of the interpolated isochrone, assuming Gaussian distributions with widths 0.01~mag in magnitude and 0.02~mag in colour, and take the highest likelihood point as the best match. This also provides, for each star, a probability that the star was drawn from the isochrone; the product of these probabilities gives the total likelihood for the high-quality sample for the given the isochrone.

We fix the adopted ages, metallicities, distance moduli and extinctions, sometimes rounding to the nearest value for which an isochrone is available. Due to the offset between the isochrone availability and the true cluster values, and the uncertainty on the cluster properties, the isochrones do not always give very good fits to the data. We have also neglected the possibility of multiple stellar populations; properly accounting for their different He and light-element abundances would also lead to slightly different masses\footnote{Although this effect is likely to be small, as shown in a study of NGC\,6352 by \citet{Libralato2019}.}. Full isochrone fitting, interpolating isochrones between grid values, and a proper accounting of multiple stellar populations are beyond the scope of this paper. Instead, to account for these sometimes-poor fits and compensate for uncertainties in the fixed quantities, we determine shifts in magnitude ($\Delta_\mathrm{m}$) and colour ($\Delta_\mathrm{c}$) that make the adopted isochrone best fit the high-quality sample. We assume flat priors on $\Delta_\mathrm{m}$ and $\Delta_\mathrm{c}$ so that the total posterior probability of the shifted isochrone given the data is equal to the likelihood of the data given the shifted isochrone and then find the offsets that maximise the total posterior probability.

Once the best shifts have been found, we once again identify the closest point on the shifted isochrone but this time for all stars (i.e. the full sample) and adopt the mass of that point as the stellar mass of the star.

This method gives us reasonable mass estimates for stars on the main sequence (MS), subgiant branch (SGB) and red giant branch (RGB), but is unreliable for horizontal branch (HB) stars and white dwarfs (WDs). To account for this, we identify regions where we observe HB and WD stars and reassign their masses. We know that significant mass loss can occur along the RGB, but the timescale for this mass loss is so fast, much faster than the typical 2-body relaxation time, that these stars will still preserve the same kinematics they had before the mass loss. So we assign all HB stars to have masses equal to the maximum in the sample, representative of their pre-mass-loss mass that still defines their kinematics. WDs are more evolved and will have experienced further mass loss and over longer, more-meaningful timescales. We adopt a mass of $0.54$~\Msun for our WD samples based on WD masses for typical old, metal-poor populations drawn from the Modules for Experiments in Stellar Astrophysics (MESA) Isochrones and Stellar Tracks \citep[MIST][]{Choi2016} models.

\begin{figure}
    \centering
    \includegraphics[width=\linewidth]{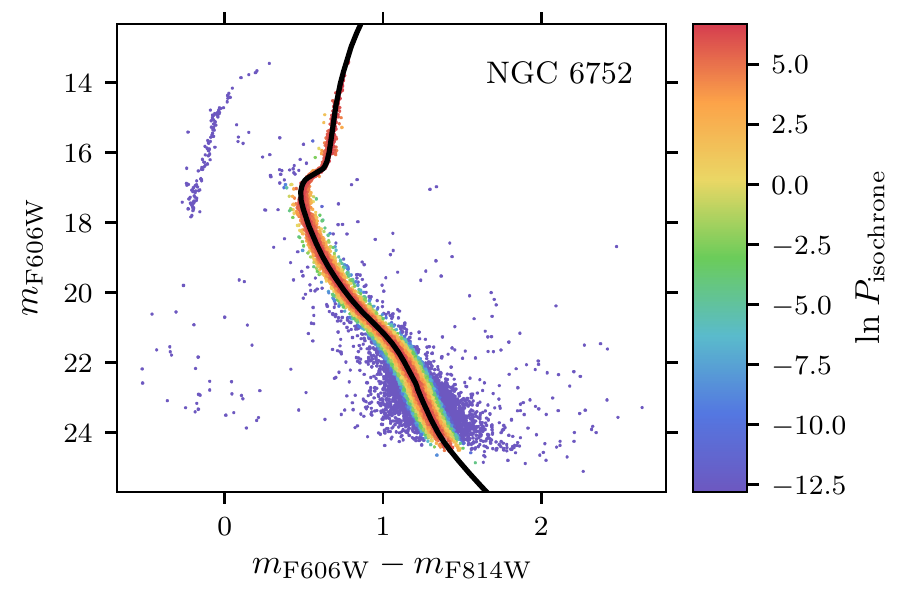}
    \includegraphics[width=\linewidth]{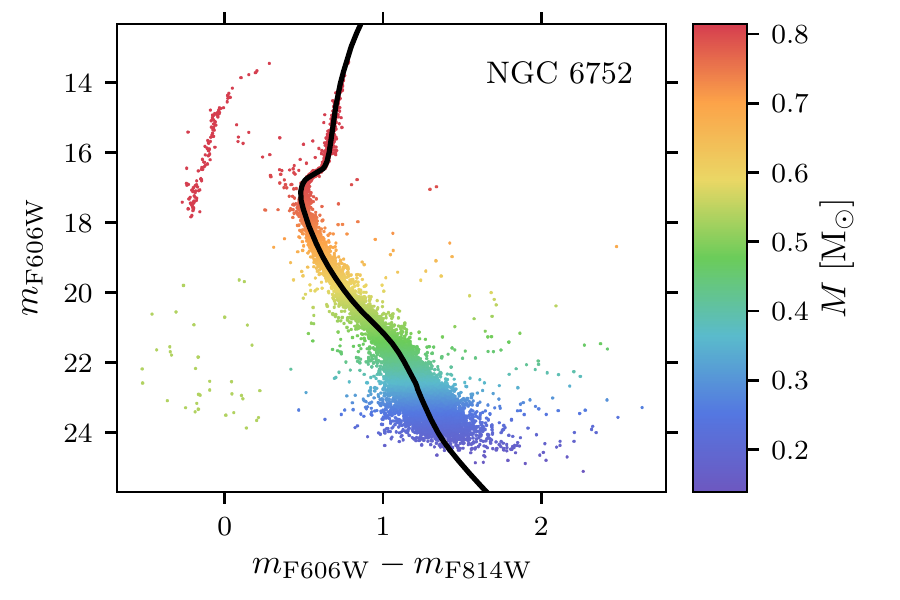}
    \caption{Colour-magnitude diagrams for NGC\,6752. Here we show the full catalogue with no quality selections or velocity clipping. Top: The data points are coloured by proximity to the isochrone, as indicated by the colour bar. Bottom: The data points are coloured by their estimated stellar mass, as indicated by the colour bar. In both panels, the black line shows the adopted isochrone for the cluster.}
    \label{figure:CMDs}
\end{figure}

As an example, \autoref{figure:CMDs} shows the colour-magnitude diagram for NGC\,6752 for all stars in the catalogue. We will use this cluster as a reference throughout the paper. The black lines show the adopted isochrone for the cluster, shifted so as to best fit the data. In the upper panel, the stars are coloured by their individual probability of being drawn from the isochrone, as indicated by the colour bar. In the lower panel, the stars are coloured by their estimated stellar mass, as indicated by the colour bar.

\section{Data Cleaning}
\label{section:datacleaning}

Before we are able to carry out the kinematic analysis, we must clean the catalogues of contaminants and ensure that we have samples of stars with well-measured velocities and well-estimated uncertainties. Including contaminants, poor measurements, or underestimated uncertainties will tend to bias our results towards larger dispersions.

In \citetalias{Watkins2015a}, we described the cleaning of a bright-star sample in detail; for these samples we were able to ignore the effect of stellar mass on any of the cluster properties. In this work, we are using all stars in the catalogue, including those on the main sequence, which significantly increases the range of stellar mass covered, so our previous cleaning method is inadequate. Here we describe our modified cleaning method. The overall process follows that from \citetalias{Watkins2015a}; the main change is to account for variations in stellar mass where appropriate, although there are a few updates to other aspects.

\subsection{PM Fit Quality}
\label{section:pmfitquality}

These cuts assess the quality of the PM fits and are similar to those used for our bright-star samples. We briefly summarise them here but refer to \citetalias{Watkins2015a} for in-depth discussion.

To calculate PMs, an initial fit was performed using all the measurements of positions versus time found for a given star $\Nfound$. Outlier points were removed and the remaining points were refit in an iterative process until the fits converged, with the final fit using $\Nused$ points. We wish to keep only stars for which the number of points rejected for the fit was low, and so remove stars with $\Nused/\Nfound < 0.85$.

The catalogues also provide reduced $\chi^2$ values separately for the fits in the $x$ and $y$ directions. For each direction, we calculate $\alpha$, the cumulative distribution function for a $\chi^2$ distribution with $D = \Nused - 2$ degrees of freedom. We remove stars for which $(1 - \alpha_\mathrm{x}) (1 - \alpha_\mathrm{y}) < 0.05$.

\subsection{Position Fit Quality}
\label{section:positionfitquality}

These cuts assess the quality of point-spread function (PSF) fits to the stars that were used to determine their positions. The principles behind these cuts are the same as for the bright-star sample, but we have refined the method, as we will describe.

Accurate PM measurement relies on accurate measurement of the positions of the stars. Blending -- where the light from a star overlaps with its neighbour, particularly a problem in crowded regions -- will cause poor PSF fits, and thus lead to large uncertainties on position. The quality of the PSF fit is given in the catalogues as a \textsc{qfit} parameter (for all clusters except $\omega$\,Centauri, for which an rms is given), for which small values indicate better fits. The quality of the PSF fits varies with magnitude as PSFs can be better fit for brighter stars, so these cuts must be magnitude dependent. We must also consider that the clusters will be more crowded (and thus subject to poorer PSF fits) in the central regions than in the outer parts, where stars are more isolated.

For NGC\,5139, we use only the F658N magnitude and rms as it is the only filter with an rms for every star in the catalogue. For the remaining clusters, we use magnitudes and \textsc{qfit} values for all available filters (F390W and F658N for NGC\,6266, or F606W and F814W for the rest).

For each star, we wish to compare its \textsc{qfit} value to \textit{isolated} stars of similar magnitude. To identify isolated stars, we bin the stars into pixels in $(x,y)$ position using a pixel scale of 10 arcsec/pixel (or 5 arcsec/pixel for NGC\,7099 due to the limited radial coverage and very low number of stars in the catalogue). We then count the number of stars in each pixel and calculate the catalogue surface density $\Sigma_\star$ for each pixel. For some clusters, the density maps are reasonably symmetric, and a simple radial cut would work well. However, the spatial and temporal coverage of the datasets that have been compiled for each catalogue are sometimes highly heterogeneous, and so the catalogue surface density of stars is highly asymmetric for some clusters.

\begin{figure}
    \centering
    \includegraphics[width=0.8\linewidth]{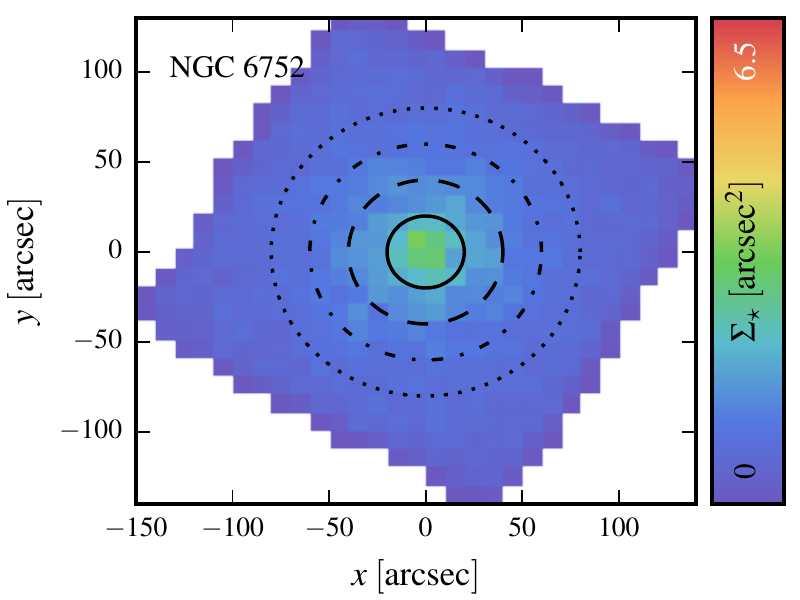}
    \includegraphics[width=0.8\linewidth]{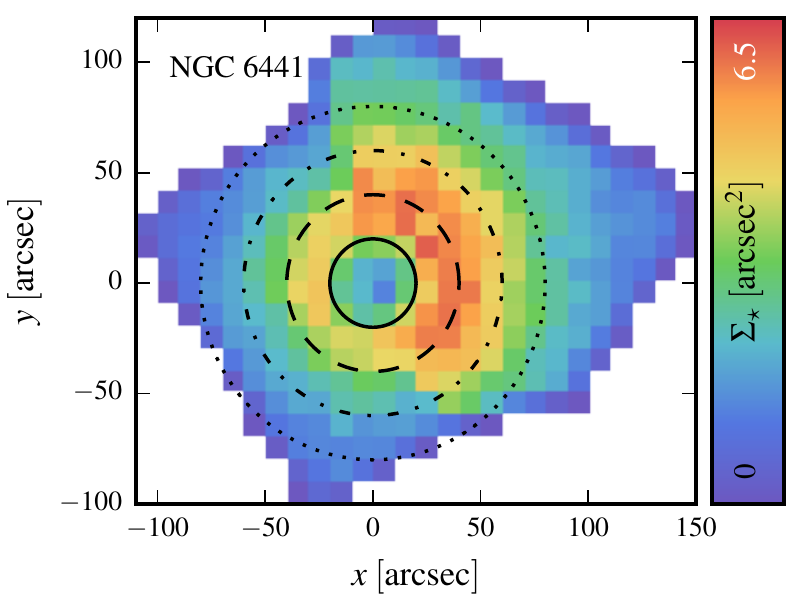}
    \caption{Surface density profiles of catalogued stars for NGC\,6752 (top) and NGC\,6441 (bottom). The colour of the pixels shows the surface density of stars in the pixel; the colour scale is the same in plots so they may be directly compared. The solid, dashed, dot-dashed, and dotted circles mark 20, 40, 60, 80~arcsec from the centre.}
    \label{figure:surfacedensity}
\end{figure}

As an example, \autoref{figure:surfacedensity} shows the catalogue surface density maps for NGC\,6752 and NGC\,6441. The colour of the pixels shows the surface density and both clusters have been plotted on the same colour scale so that they may be directly compared. The circles indicate 20 (solid), 40 (dashed), 60 (dot-dashed), and 80~arcsec (dotted). For NGC\,6752, the surface density is fairly symmetric about the centre. However, for NGC\,6441, the density is asymmetric due to the off-axis field towards the top-right of the figure.

Note also that the catalogue surface density for NGC\,6752 is fairly low everywhere -- primarily because it is one of the closest clusters in our sample -- and increases towards the centre. For NGC\,6441, we see a central dip in the density map; this is not because the density drops, but because the incompleteness is so high here that many stars in this region did not make it into the PM catalogues. We need to account for both the asymmetry in the catalogue density maps and the central holes when selecting isolated stars. We experimented with different definitions and find that defining isolated stars as those where $\Sigma_\star < 2.5$~stars/arcsec$^2$ works well for our purposes. For the 6 clusters with central holes (NGC\,1851, NGC\,2808, NGC\,6388, NGC\,6441, NGC\,6715, and NGC\,7078), we further insist that isolated stars must lie outside of 20~arcsec.

For our \textsc{qfit} analysis, we begin by assigning a weight to all other stars in the catalogue based on the proximity in magnitude, for which we use a Gaussian centred on the target star with width 0.1~mag. Then to ensure that we are comparing only against isolated stars, we set the weight of all stars not selected as part of the isolated sample to be 0.

We then calculate a normalised weighted cumulative distribution function of \textsc{qfit} values, and then use this to determine the (weighted) percentile $\xi_Q$ in which the target star lies. In our previous analysis, we also calculated \textsc{qfit} percentiles, but using only the nearest 100 stars in magnitude outside of some radial limit, instead of weighting the contribution of the stars by their proximity in magnitude and performing a density analysis.

To select high-quality stars, we remove all stars with $\xi_Q > 65\%$. For NGC\,5139, we use only the F658N filter for the cut as the rms statistics for the other bands are incomplete, as discussed above. For all other clusters, we use \textsc{qfit} statistics for both filters available, and remove a star that fails the cut in either filter.

\begin{figure*}
    \centering
    \includegraphics[width=\linewidth]{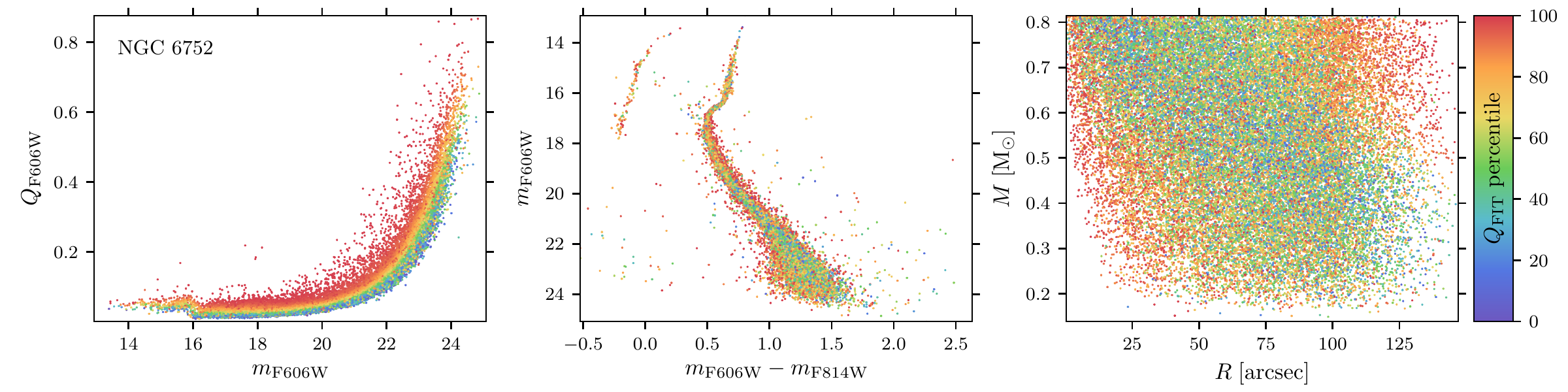}
    \caption{PSF fit quality statistics for NGC\,6752. Left panel: \textsc{qfit} as a function of magnitude for F606W. The equivalent plot for F814W is very similar and so is not shown. Middle panel: colour-magnitude diagram. Right panel: location of the stars in the plane of distance from the cluster centre and stellar mass. In all three panels, the colour of the points shows the \textsc{qfit} percentile calculated as described in the text and shown in the colour bar.}
    \label{figure:qfits}
\end{figure*}

\autoref{figure:qfits} demonstrates the \textsc{qfit} percentiles for NGC~6752. This is one of the closer clusters in our sample, so the \textsc{qfit} values are generally well behaved. In the left panel, we show the \textsc{qfit} for the F606W photometry, as a function of magnitude. We see that the PSF is fit better for bright stars and becomes worse for fainter stars. The points are coloured by the percentile they were assigned based on their \textsc{qfit}, as shown in the colour bar on the right. We have two percentiles (one for F606W and one for F814W) and the colours show whichever is the larger value. The plot for the F814W is very similar so we do not show it here.

The middle panel shows a colour-magnitude diagram for the cluster, with the colours the same as before. We see that stars near to the central locus of the population typically have better PSF fits than those at the edges, which is not surprising as photometric precision decreases (and measurement errors increase) for fainter stars.

In the right panel, we show the catalogue in the plane of distance from the cluster centre and stellar mass, again coloured by \textsc{qfit} percentile as shown by the colour bar. We see that, unsurprisingly, faint stars near the crowded centre are typically poorly measured and will be removed.

\subsection{Velocity Uncertainties}
\label{section:velocityuncertainties}

These cuts consider the reliability of the uncertainties on the PM measurements. In \citetalias{Watkins2015a}, we folded this into the analysis of the local dispersions (the next step, which we will discuss in \autoref{section:localdispersions}). Here we separate the two analyses.

The observed velocity dispersion is the convolution of the intrinsic dispersion of the stars and their velocity uncertainties. So to recover the true dispersion, we need accurate measurement uncertainties. Underestimated (overestimated) uncertainties will lead to an estimate of intrinsic dispersion that is higher (lower) than the true value. In general, uncertainties can be over- or underestimated. In practice for this type of work, well measured stars tend to have accurate uncertainties, but poorly measured stars tend to have underestimated uncertainties.\footnote{This is because uncertainty estimations generally assume that the PM is well measured; when the PM is not well-measured, the uncertainty estimation becomes unreliable.} So smaller uncertainties are generally reliable, whereas larger uncertainties, although large, tend to be underestimated and are not large enough. Thus, we will want to keep smaller uncertainties and remove larger ones.

Here we are not concerned with the overall size of the uncertainties (that will come later), just whether they have been accurately estimated, which we will assess by comparing to measurements for other similar stars.

Positions for bright stars can be determined more accurately than for faint stars. It follows then that the change in position with time (or PM measurement) can also be determined more accurately for bright stars than for faint stars. So PM uncertainty changes as a function of magnitude, and this needs to be factored in to our analysis of reliability.

PM uncertainties also change as a function of baseline, with longer time baselines yielding more precise PM measurements. Some of our PM catalogues are constructed from multiple fields with different baselines, and we see the effect of this in the uncertainties. PM uncertainties also vary depending on the number of observations used for the measurement, and with exposure time of the observations. Some of these effects will be small compared to the overall scatter in the uncertainties, but some will be meaningful. To assess the status for each cluster, we begin by reviewing their distributions of PM uncertainty as a function of magnitude. For 16 clusters, we identify only one sequence of uncertainties, indicating that the effect of baseline, number of observations and exposure time, and any other factors is smaller than the overall scatter. In the other 6, we identify multiple sequences, indicating that various factors are dominating the distributions. For these clusters, we isolate the sequences and perform the following cleaning method on each sequence separately.

The 6 clusters, and the cuts used to isolate their sequences are:
\begin{itemize}
    \item NGC\,104, 2 sequences: T $<$ 7 yr, T $>$ 7 yr;
    \item NGC\,5139, 2 sequences: T $<$ 7.5 yr, T $>$ 7.5 yr;
    \item NGC\,6388, 3 sequences: T $<$ 3 yr, 3 yr $<$ T $<$ 9 yr, T $>$ 9 yr;
    \item NGC\,6397, 2 sequences: T $<$ 4 yr, T $>$ 4 yr;
    \item NGC\,6441, 3 sequences: T $<$ 6.5 yr, T $>$ 6.5 yr and $\Nused < 18.5$, T $>$ 6.5 yr and $\Nused > 18.5$;
    \item NGC\,7078, 2 sequences: T $<$ 5 yr, T $>$ 5 yr;
\end{itemize}
for baseline $T$ and number of observations used for the PM measurement $\Nused$.

To proceed, we follow a similar approach to that outlined in \autoref{section:positionfitquality} for making \textsc{qfit} cuts. We first identify a well-measured sample, which we define as stars where the local catalogue density is $0.5 < \Sigma_\star < 2.5$~stars/arcsec$^2$ and the \textsc{qfit} percentile $\xi_Q < 50\%$. This is a stronger \textsc{qfit} cut than we used in \autoref{section:positionfitquality} to make sure that we are really taking the best measured stars for this sample. Then, for each star, we weight the well-measured stars according to their proximity in magnitude using a Gaussian centred on the target star of width 0.1~mag, calculate the normalised weighted cumulative distribution function of uncertainties $\Delta_\mu$, and determine the weighted percentile $\xi_\Delta$ in which the star lies. We do this separately for the uncertainties in the x and y directions and use the maximum of these values for each star. To select high-quality stars, we remove all stars with $\xi_\Delta > 65\%$.

\begin{figure*}
    \centering
    \includegraphics[width=\linewidth]{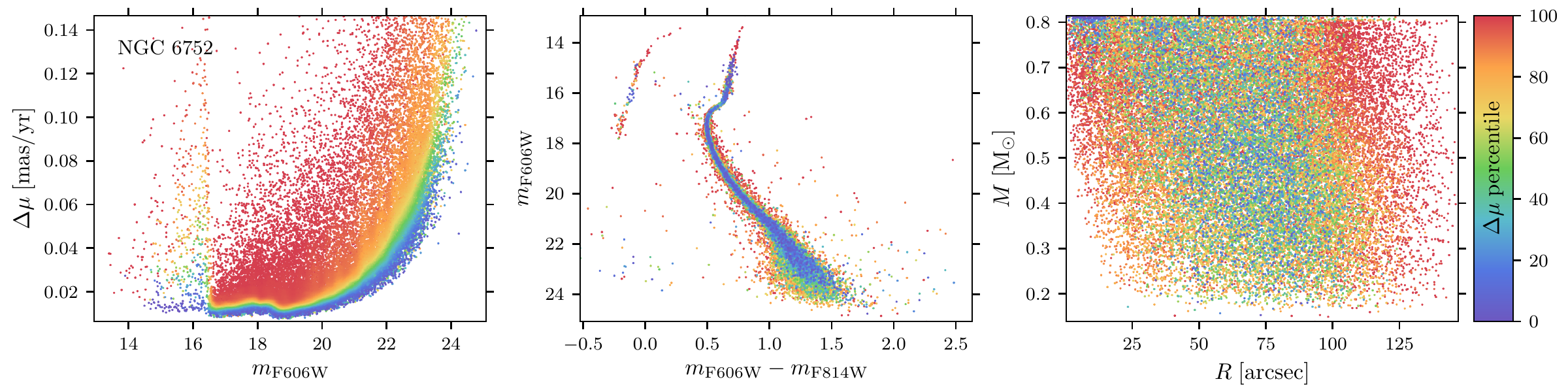}
    \includegraphics[width=\linewidth]{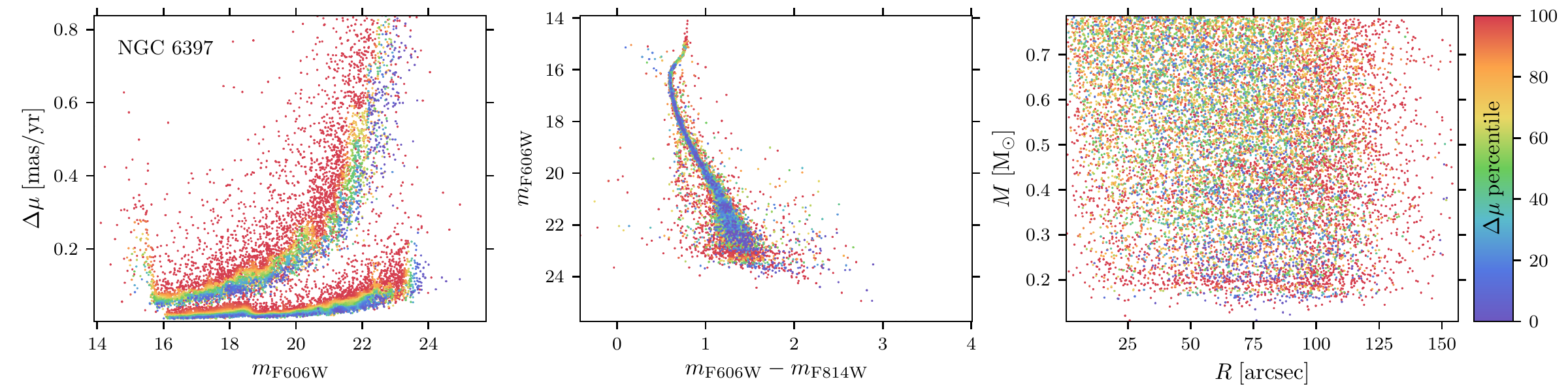}
    \caption{PM uncertainty quality statistics for NGC\,6752 (top, our reference cluster in this work) and NGC\,6397 (bottom, an example with two sequences). Left panels: PM uncertainty $\Delta_\mu$ as a function of magnitude for F606W. Uncertainties in both $x$ and $y$ are shown in this panel. The equivalent plot for F814W is very similar and so is not shown. In the upper panel there is only one sequence. In the lower panel, two sequences are clearly identified, isolated with a cut at baseline 7~yr for this cluster. The quality analysis was run separately for both sequences. Middle panels: colour-magnitude diagram. Right panels: location of the stars in the plane of distance from the cluster centre and stellar mass. In all three panels, the colour of the points shows the percentile $\xi_\Delta$ calculated as described in the text and shown in the colour bar.}
    \label{figure:pmerrors}
\end{figure*}

\autoref{figure:pmerrors} demonstrates the PM uncertainty percentiles for NGC\,6752 (upper panels) and NGC\,6397 (lower panels). These are both among the closer clusters in our sample, so the uncertainties are generally well behaved. In the left panel, we show the PM uncertainties (in both $x$ and $y$ directions) as a function of magnitude. We see that the uncertainties are smaller for the brighter stars, for which positions can be determined more accurately, than for the fainter stars. For NGC\,6752, we see only one sequence. For NGC\,6397, two sequences are evident due to different baselines for PMs in this catalogue. The points are coloured by the percentile they were assigned based on their PM uncertainty, as shown in the colour bar on the right. We have two percentiles (one for the $x$ motions and one for $y$ motions) and the colours show whichever is the larger value.

The middle panel shows a colour-magnitude diagram for the cluster, with the colours the same as before. As for the \textsc{qfit} analysis, we see that stars near to the central locus of the population typically have better PM uncertainties than those at the edges, which is not surprising if we consider that stars near the locus were likely better measured.

In the right panel, we show the catalogue in the plane of distance from the cluster centre and stellar mass, again coloured by percentile $\xi_\Delta$ as shown by the colour bar. The patterns in these plots change from cluster to cluster, owing to differences in the intrinsic cluster properties and in the observations used for these catalogues. But in general, stars near the centre tend to have large percentiles (large uncertainties relative to other similar stars, coloured red) due to crowding and small-number statistics. Stars in the outer parts tend to have large percentiles due to less accurate geometric distortion solutions and/or PSFs near the field edges, though small-number statistics may have an effect here as well.

\subsection{Local Velocity Dispersions}
\label{section:localdispersions}

These cuts are designed to remove contaminant populations with velocity distributions offset from that of the cluster and to remove stars with uncertainties much larger than the signal we are trying to measure. Both the velocity and uncertainty cuts depend on the velocity dispersion, which changes as a function of both radius and stellar mass. \citetalias{Watkins2015a} considered only changes in dispersion as a function of radius, so we have to update this step to fold in mass dependence for the dispersions. As we require a clean sample of stars to calculate reliable kinematics, this is necessarily an iterative process.

We begin with a crude membership selection via sigma-clipping to identify extreme outliers. To do this, we assume that the mean velocity of the sample is zero (as it should be by design as the catalogues provide only relative PMs) and calculate the rms of the radial and tangential PMs, which is the dispersion $\sigma$ under the assumption of zero mean. We then flag as `good' all stars within 5$\sigma$ of zero and for which the PM uncertainties are smaller than 2$\sigma$. We then recalculate the dispersion using the `good' stars and reassign `good' flags based on the new results. We iterate until the result is stable. In this step, we do not consider mass or radial dependence on the velocity dispersion, neither do we include the uncertainties on the velocity measurements in the dispersion estimation, so the dispersion we calculate is an average of the intrinsic cluster dispersion convolved with the measurement uncertainties; at this point, all we want to do is identify obvious outliers (e.g. foreground stars), so this crude clipping is good enough. We do not remove any stars here, we only flag them as `bad' for the initial step of the next phase.

Now, we have a roughly cleaned sample, we are ready to calculate local dispersions based on mass and radius. We assume a dispersion model that behaves like a 3rd-order polynomial as a function of radius $R$ and power-law as a function of mass $m$, that is
\begin{equation}
	\sigma \left( R, m \right) = \left[ c_0 + c_1 R + c_2 R^2 + c_3 R^3 \right] m^{-c_4} .
\end{equation}
$c_4$ is analogous to the traditional equipartition parameter, but we refrain from calling it such because here we are not fitting for equipartition, just cleaning.

As a baseline, we assume flat priors on all the parameters, but in some cases control their allowed range. The polynomial part is essentially the dispersion profile of a 1~\Msun star, so we insist that is must be positive for all data points, that is,
\begin{equation}
	c_0 + c_1 R_i + c_2 R_i^2 + c_3 R_i^3 > 0,
\end{equation}
for all stars $i$ with radii $R_i$. We also force the polynomial part to be monotonically decreasing, as we would expect for a dispersion profile, that is,
\begin{equation}
	c_1 + 2 c_2 R_i + 3 c_3 R_i^2 > 0,
\end{equation}
for all radii $R_i$. Finally, we insist that $-5 \le c_4 \le 5$. A physically motivated choice would restrict $0 \le c_4 \le 0.5$, but this does not account for the possibility that the cleaned catalogues may not be of sufficient quality to perform this kind of analysis. So we choose a looser prior that will keep the models from straying too far away from the expected values, but does not force them to be in the expected range. Indeed, if $c_4$ is outside of the expected range (and especially if it hits these boundaries), it is a clear sign that the data quality is not sufficient for this work for that cluster at present.

We fit this model to the high quality, good stars (with high quality defined by the PM and position fit quality cuts and good defined by the previous sigma-clipping). We perform the fits using a maximum-posterior analysis in a discrete fashion, that is the fit is assessed using individual stars not binned kinematics. For a star $i$ at position $\Ri$ with mass $\mi$ and velocity $( \mu_\mathrm{R,i} \pm \delta_\mathrm{R,i}, \mu_\mathrm{T,i} \pm \delta_\mathrm{T,i})$, the model predicts a mean 0 and dispersion $\sigma_\mathrm{i} = \sigma(\Ri,\mi)$. The total likelihood is the product of the individual likelihoods for each star and is then
\begin{equation}
	\mathcal{L} = \prod_\mathrm{i} \frac{1}{2 \pi \sigma_\mathrm{i}^2} \exp \left[ - \left( \frac{\mu_\mathrm{R,i}^2}{\left( \sigma_\mathrm{i}^2 + \delta_\mathrm{R,i}^2 \right)} + \frac{\mu_\mathrm{T,i}^2}{\left( \sigma_\mathrm{i}^2 + \delta_\mathrm{T,i}^2 \right)} \right) \right] .
	\label{equation:likelihood}
\end{equation}
Finally, the posterior is the product of the priors and the likelihood. We maximise the posterior using a downhill simplex method.

Using the best-fitting dispersion predicted for each star $\sigma_\mathrm{i}$, we once again sigma-clip the sample, but now using the best-fitting model dispersion convolved with the observational uncertainties. That is, we compute
\begin{equation}
	N_\mathrm{\sigma,i} = \left( \frac{\mu_\mathrm{R,i}^2}{\sigma_\mathrm{i}^2 + \delta_\mathrm{R,i}^2} + \frac{\mu_\mathrm{T,i}^2}{\sigma_\mathrm{i}^2 + \delta_\mathrm{T,i}^2} \right)^\frac{1}{2}
\end{equation}
and flag as good all stars with $N_\mathrm{\sigma,i} \le 3$.

To measure accurate dispersions, it is important to have measurements for which the uncertainties are well below the internal motions of the cluster \citepalias[see broader discussion in][]{Watkins2015a}. So we flag as `bad' any star for which $\delta_\mathrm{R,i} > \frac{1}{4} \sigma_\mathrm{i}$ or $\delta_\mathrm{T,i} > \frac{1}{4} \sigma_\mathrm{i}$. In general, the uncertainties increase faster with increasing magnitude (due to limited signal-to-noise) than do the internal motions (due to equipartition). Therefore, this step mostly eliminates faint stars, and effectively determines a the lowest stellar mass down to which we can measure the cluster dispersion.

As before, we then iterate using the new good flags to re-fit the dispersion model and then recalculate the good flags until the good sample converges, or until 30 iterations have been performed.

\begin{figure*}
    \centering
    \includegraphics[width=0.95\linewidth]{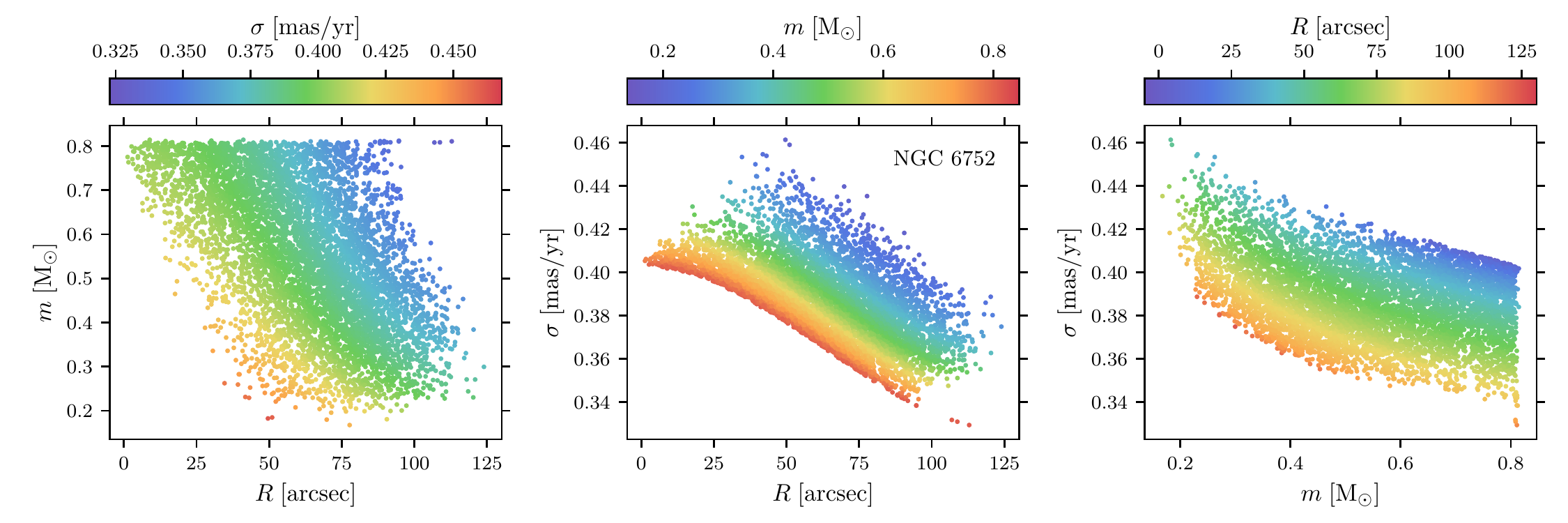}
    \caption{Local dispersion cleaning for NGC\,6752. Left panel: location of the final sample in the plane of distance from the cluster centre and stellar mass, coloured by the local dispersion estimated from the fit as indicated by the colour bar above. Middle panel: fitted velocity dispersion as a function of distance from the cluster centre for the final sample, coloured by the stellar mass as indicated by the colour bar above. Right panel: fitted velocity dispersion as a function of stellar mass for the final sample, coloured by the distance from the cluster centre as indicated by the colour bar above.}
    \label{figure:localdispersions}
\end{figure*}

\autoref{figure:localdispersions} shows the results of the cleaning process for NGC\,6752. The left panel shows the final sample of stars in the plane of distance from the cluster centre $R$ and stellar mass $m$, with the stars coloured by the local dispersion $\sigma$ estimated from the fit (red high and blue low) as indicated by the colour bar above. As expected, the low mass stars near the centre of the cluster move fastest, with velocity dispersion decreasing with both increasing mass and increasing distance from the centre. It is immediately apparent that the radial coverage is not the same for every mass, and the mass range is not the same at every radius.

The middle panel shows velocity dispersion as a function of distance from the cluster centre, with stars coloured by their mass as indicated by the colour bar above. We see that the high mass (red) stars have lower dispersions than the low mass (blue) stars. The right panel shows velocity dispersion as a function of stellar mass, with stars coloured by the distance from the cluster centre. The stars near the centre (blue) move faster than the stars towards the edge of the field (red). The three panels provide different views of the same data. We include all three here to aid understanding of the processes at work and the underlying distributions in these clusters.

Solutions failed for 6 clusters in the sample. For NGC\,288, NGC\,6535, and NGC\,7099, this cleaning step returned 0 stars. For NGC\,6362, this step returned just 7 stars, which is not enough to do anything meaningful. For NGC\,6624, the solution did not converge. For NGC\,6715, the solution hit the upper boundary of the $\eta$ prior. For these clusters, we conclude that the data quality is not sufficient for us to proceed with this analysis. None of these clusters are particularly surprising as they include very low dispersion clusters (NGC\,288, NGC\,6362, NGC\,6535), small sample-size clusters (NGC\,288, NGC\,6362, NGC\,6535, NGC\,6624, NGC\,7099), and the most distant cluster in the sample (NGC\,6715). NGC\,6715 may also suffer from contamination from the Sagittarius dwarf galaxy. All of these factors will make any analysis challenging.

\subsection{Final Cluster Sample}
\label{section:finalsample}

In the previous section, we calculated the local velocity dispersion $\sigma$ as a function of mass and radius, removed outliers by removing stars more than 3$\sigma$ away from 0, and improved the signal-to-noise of the sample by removing stars with PM uncertainties larger than $\frac{1}{4} \sigma$.

The latter step was extremely conservative because the signal we are looking for in the data is small, but ideally the final results should be fairly insensitive to exactly where we make this cut. Or to re-frame the point, we would like to make sure that enough stars are far enough below the cut value that its exact position does not matter. To that end, for the remaining 16 clusters, we ask for what fraction of stars is $\delta_\mathrm{\{R,T\},i} < \frac{1}{8} \sigma_\mathrm{i}$ and insist that this must be $0.2$. Recall the cut we made previously was at $\frac{1}{4}$, so we are insisting that at least 20\% of the stars must have errors lower than half of the cut value.

There is one final consideration, which is that we are aiming to investigate changes in velocity dispersion as a function of stellar mass. We want to ensure we have a reasonable range of stellar masses to study. The previous cuts have tended to excise low-mass stars, so the lower boundary of the mass scale can be fairly ragged in some clusters, so instead of calculating the mass range, we calculate the range spanned by the 5th percentile to the 100th percentile in mass, and insist that this must be larger than 0.15~\Msun.

A total of 7 clusters fail these cuts: 4 fail both (NGC\,1851, NGC\,6388, NGC\,6441, and NGC\,7078), 2 fail the first but not the second (NGC\,362 and NGC\,6681), and 1 fails the second but not the first (NGC\,5927).

This leaves us with a sample of 9 clusters that pass both cuts: NGC\,104, NGC\,2808, NGC\,5139, NGC\,5904, NGC\,6266, NGC\,6341, NGC\,6397, NGC\,6656, and NGC\,6752. The number of stars in the final catalogue for each cluster is given in \autoref{table:summary}.

\section{Energy Equipartition}
\label{section:equipartition}

Now that we have cleaned high-quality catalogues of PMs, and stellar mass estimates for all the stars, we can investigate how the velocity dispersion changes as a function of stellar mass. We will start off by looking at the properties of the catalogues as a whole. An important caveat is that these PM catalogues cover only the central regions of these clusters, so this does not give a global view of equipartition, only that within the region spanned by the data.

\subsection{Classic Power-Law Function Fits}
\label{section:classic}

We wish to study how velocity dispersion changes as a function of stellar mass in each cluster catalogue. We know that velocity dispersion also changes as a function of distance from the cluster centre. In this next section, we will look at these effects simultaneously, but at present, we want to consider only the effect of stellar mass.

If the radial distribution of stars in each stellar-mass bin was the same, then the effect of radius would be small and we should still be able to obtain a robust equipartition estimate. However, the distribution of stars with radius is likely not the same for stars of different mass. This could be due to physical effects in the cluster such as mass segregation whereby slower-moving high-mass stars tend to sink towards the centre and faster-moving low-mass stars tend to migrate outwards. But also due to observational effects, for example as cluster centres are dense and crowded, we tend to have high-quality measurements only for the brightest (high mass) stars in these regions, whereas less crowded regions will have high-quality measurements for bright and faint (high mass and low mass) stars (see the left panel of \autoref{figure:localdispersions}).

A competing consideration is that we need a sufficient number of stars so that we can estimate robust dispersions in a series of stellar-mass bins. So to mitigate the radial effects while ensuring we have enough stars, we select only the stars within the 25th and 75th radial percentiles for this analysis.

We have PMs, which provide us with velocity measurements in 2 orthogonal directions. In \citet{Watkins2015a}, we used the bright stars to perform an anisotropy analysis on all 22 clusters and found that many are isotropic over the full range of the data, and others are isotropic at their centres and become only mildly radial at the edges of the data. To simplify this analysis, we neglect the effect of anisotropy here and fit a single model to both the radial and tangential PMs (we return to this point later in \autoref{section:discussion}).

We begin with the classic model for energy equipartition given in \autoref{equation:equipartition}. Here we will use a scale mass $m_s$ of 1~\Msun, so we have two free parameters: $\sigma_1$, the velocity dispersion of a 1~\Msun star, and $\eta$, the degree of energy equipartition.

As discussed in \autoref{section:localdispersions}, a physically-motivated prior would restrict $\eta$ to physically-expected values, but does not account for the possibility of the data not being of sufficient quality (despite the extensive cleaning) to do this kind of analysis. As such, we use looser bounds of $\pm$5 and assume a flat distribution over the allowed range, that is,
\begin{equation}
	P(\eta) = \left\{ \begin{array}{cc}
    	1 & \qquad \left| \eta \right| \le 5 \\
        0 & \qquad \left| \eta \right| > 5
    \end{array} \right.
\end{equation}
For the scale dispersion, we insist the value must be positive and use a flat prior for allowed values, that is,
\begin{equation}
	P(\sigma_1) = \left\{ \begin{array}{cc}
    	1 & \qquad \sigma_1 \ge 0 \\
        0 & \qquad \sigma_1 < 0
    \end{array} \right.
\end{equation}

The likelihood $\mathcal{L}$ of the observed measurements given a particular model ($\sigma_1$, $\eta$) is given by \autoref{equation:likelihood}, where in this case $\sigma_i = \sigma(m_i | \sigma_1, \eta)$ for star $i$. Then the posterior probability $\mathcal{P}$ of a model ($\sigma_1$, $\eta$) given the observed data is given by
\begin{equation}
    \mathcal{P} = \mathcal{L} P(\eta) P(\sigma_1) .
\end{equation}
To find the region of parameter space where the posterior is maximised and, thus, with the best-fitting models, we run a Markov Chain Monte Carlo (MCMC) analysis. We use the affine-invariant MCMC package \textsc{emcee} \citep{ForemanMackey2013} with 100 walkers for 1000 steps.

\begin{figure}
    \centering
    \includegraphics[width=\linewidth]{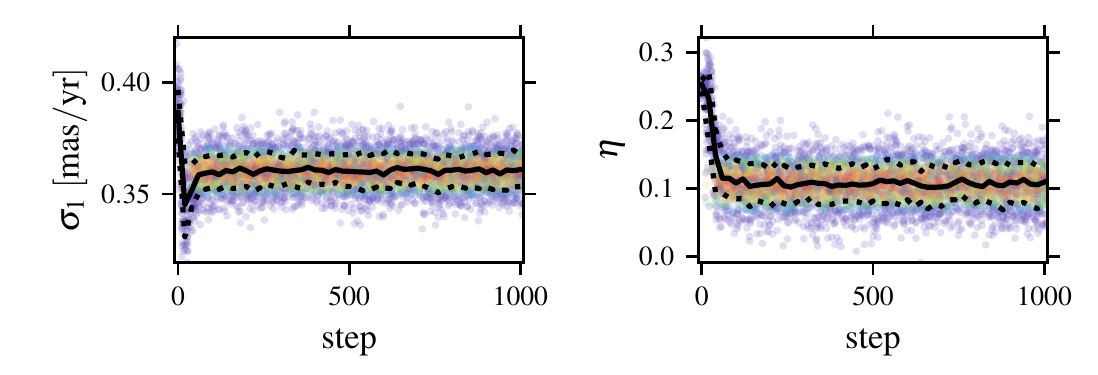}
    \includegraphics[width=0.9\linewidth]{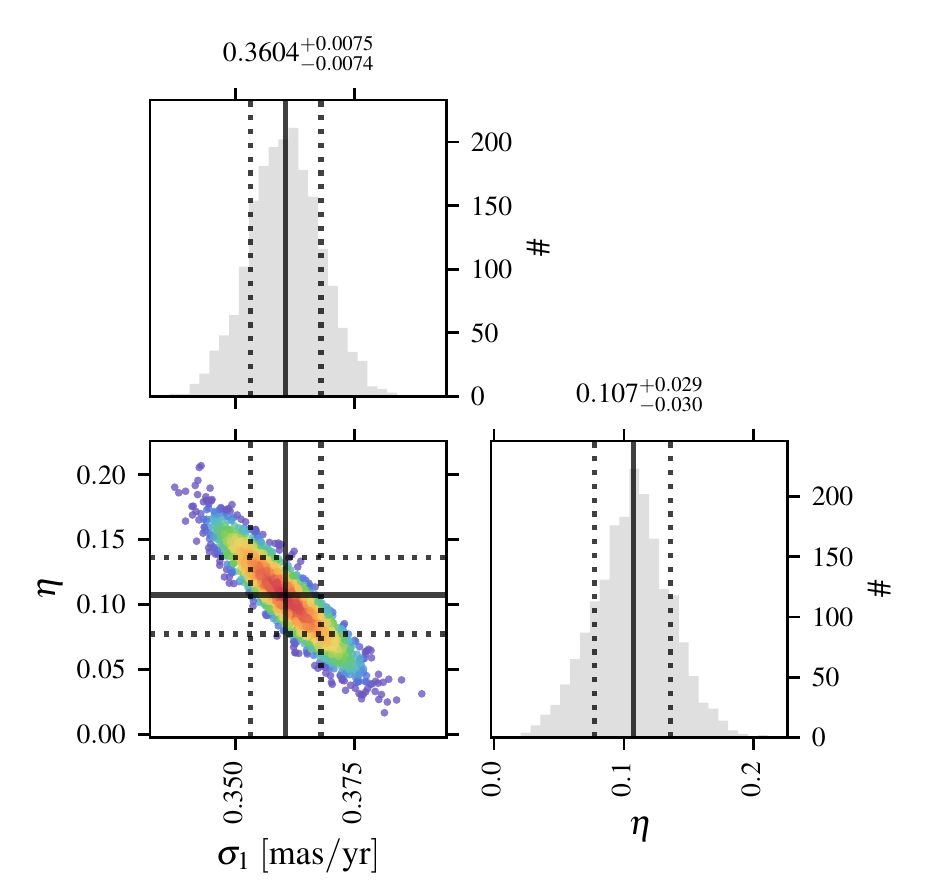}
    \caption{MCMC chain results for the 1d classic power-law fit to NGC\,6752. Top panels: progression of the two free parameters $\sigma_1$ (left) and $\eta$ (right) for the whole run. Both panels are flat showing that the chain has converged well. Bottom panels: distribution of walkers at the end of the run. The histograms show the distribution of the individual $\sigma_1$ and $\eta$ parameters and are approximately Gaussian; the scatter plot shows their correlation in phase-space.}
    \label{figure:1dClassicChain}
\end{figure}

The fits for all 9 clusters converged. As an example, in \autoref{figure:1dClassicChain} we show the location of the walkers along the whole length of the chain (upper panels) and the final distribution of the walkers (lower panels) for the fits to NGC\,6752. Colours indicate posterior probability from high (red) to low (purple). To obtain best estimates for these fitted values, we take the positions of all the walkers in 20 steps extracted at 10-step intervals from the end of the chain. This gives us 2000 points in total -- we will call these the final sample for each cluster going forwards. We adopt the median of these as the best estimate for the fitted values, and use the 15.9 and 84.1 percentiles to estimate the lower and upper uncertainties. As shown in \autoref{figure:1dClassicChain}, these distributions are approximately Gaussian so the median is approximately the same as the mean, and the upper and lower error bars are approximately equal to each other and approximately the same as the standard deviation.

To display the resulting fits, we also calculate a binned dispersion profile. We stress that the fits were performed using the individual measurements, the binned profile is for visualisation only. We subdivide the data into bins such that each bin is equally populated; the number of bins is chosen such that every bin has at least 25 stars (thus 50 PM measurements as each star contributes two PMs) up to a maximum of 10 bins. This ensures we have enough measurements per bin to estimate a statistically reliable dispersion and uncertainty but also ensures we do not have so many bins that the signal is overwhelmed by noise. In each bin, we use simple maximum-posterior analysis to estimate the velocity dispersion along with the corresponding uncertainties.

\begin{figure*}
    \centering
    \includegraphics[width=0.3\linewidth]{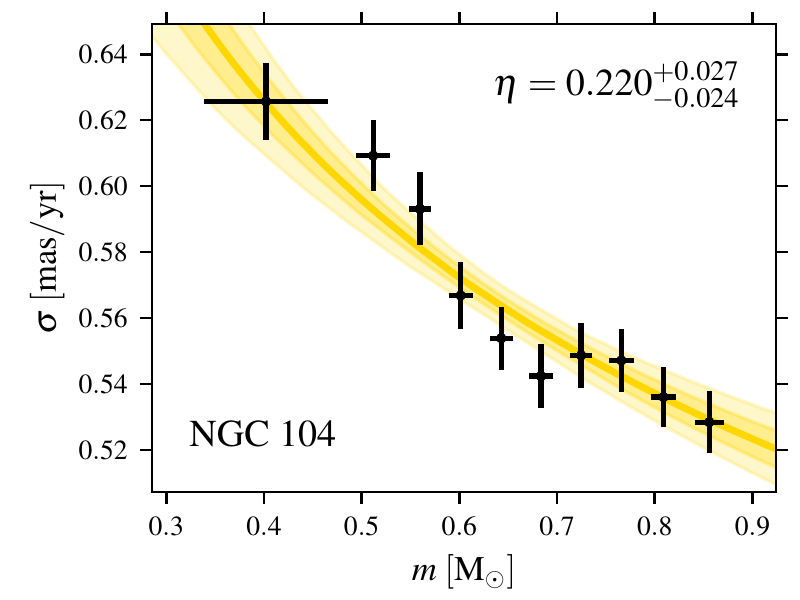}
    \includegraphics[width=0.3\linewidth]{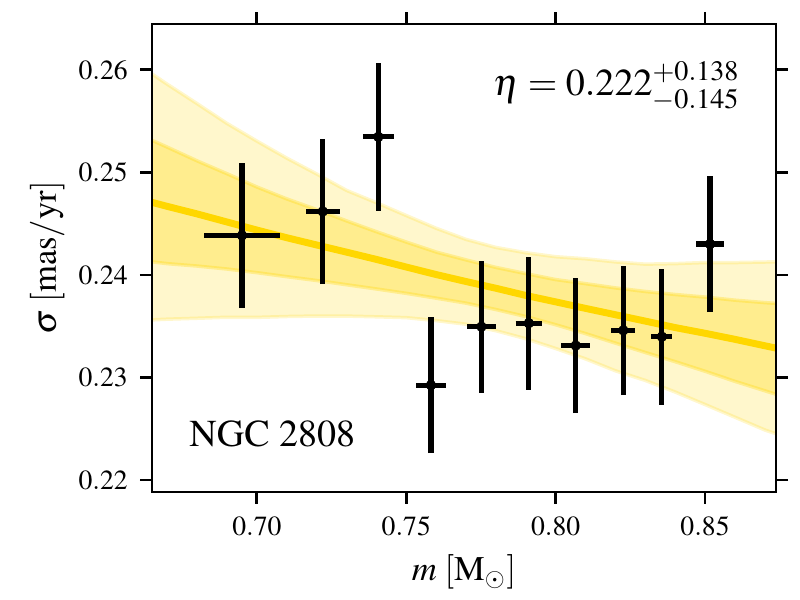}
    \includegraphics[width=0.3\linewidth]{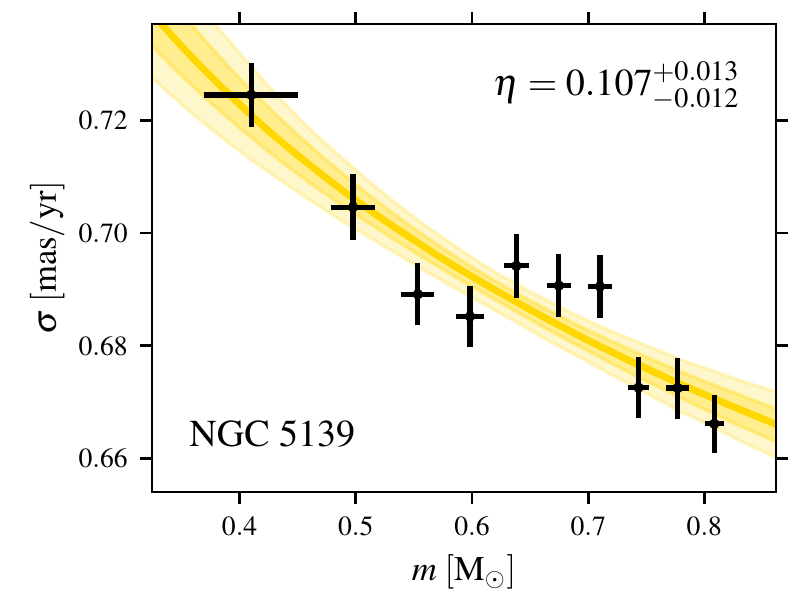}
    \includegraphics[width=0.3\linewidth]{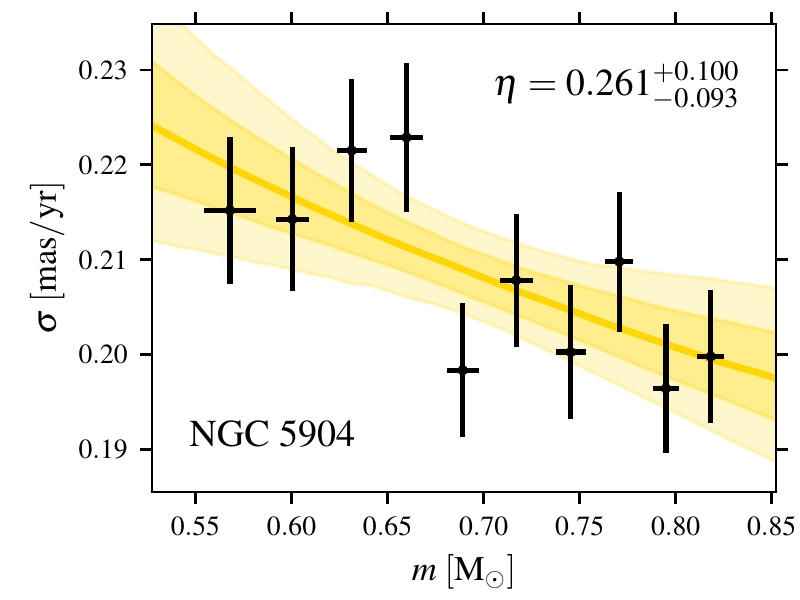}
    \includegraphics[width=0.3\linewidth]{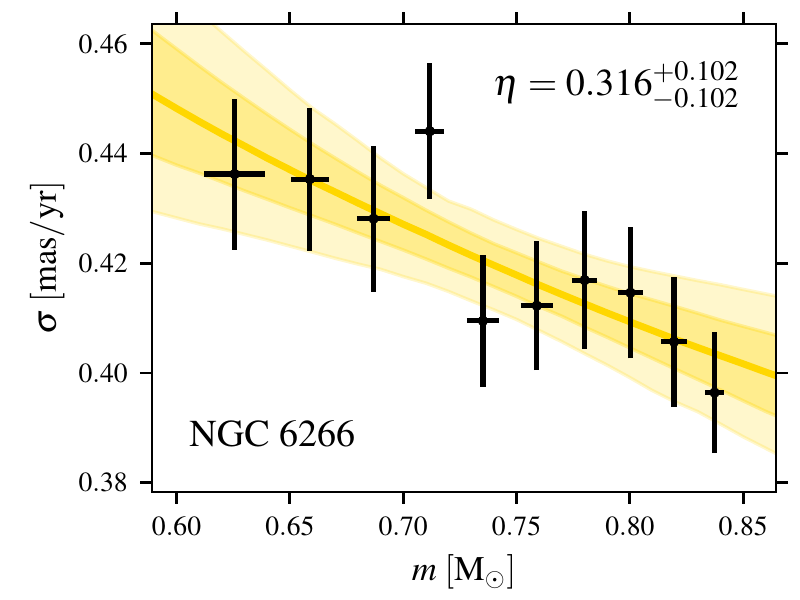}
    \includegraphics[width=0.3\linewidth]{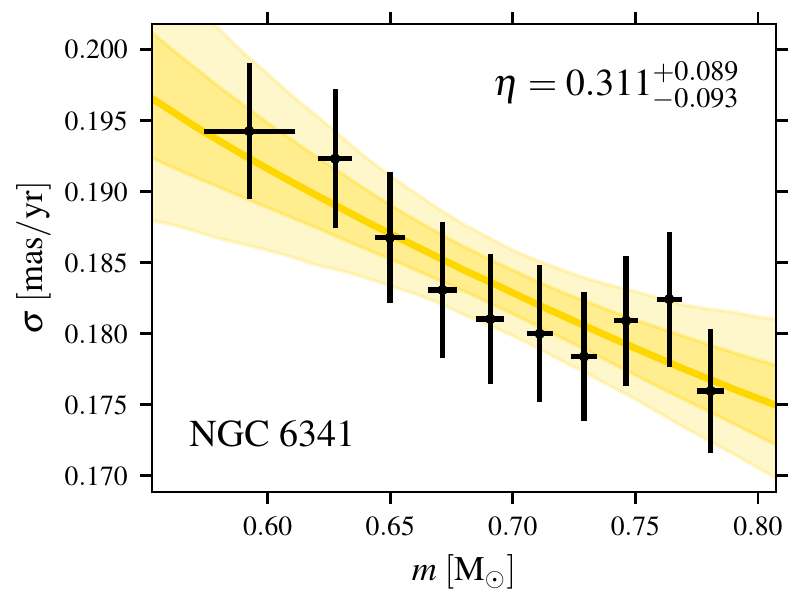}
    \includegraphics[width=0.3\linewidth]{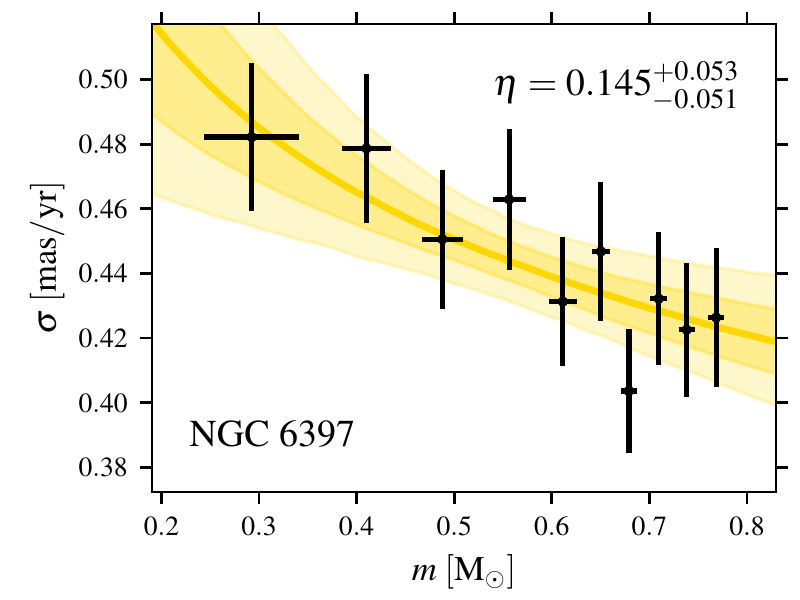}
    \includegraphics[width=0.3\linewidth]{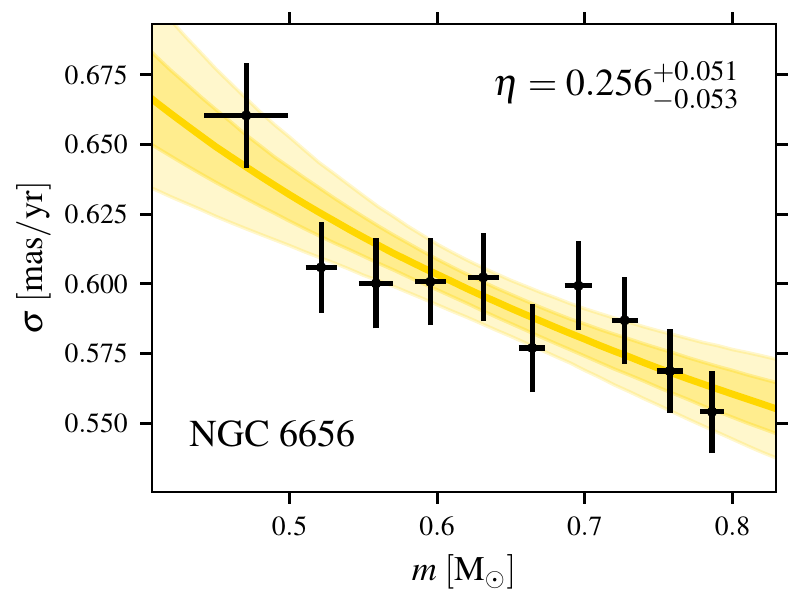}
    \includegraphics[width=0.3\linewidth]{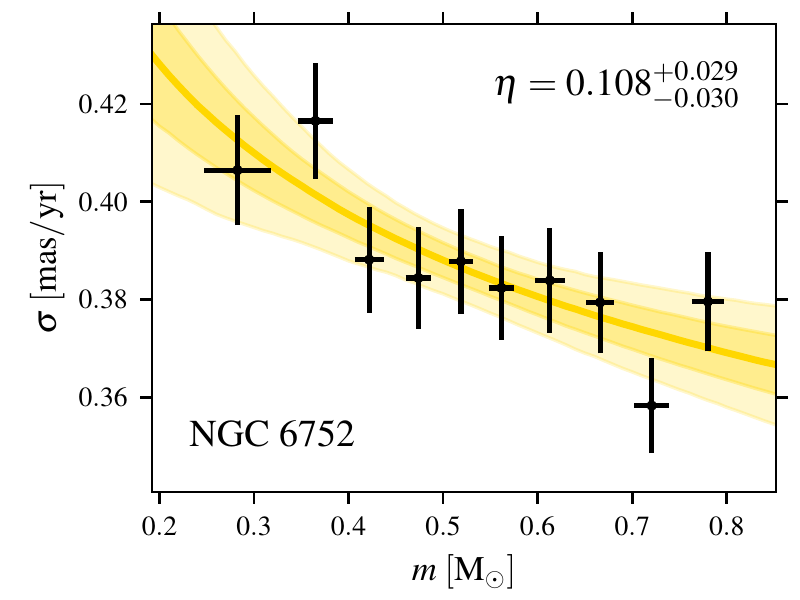}
    \caption{Power-law fits to the velocity dispersion as a function of stellar mass for the 9 clusters in our sample. The solid yellow lines show the median of the fitted profiles. The darker shaded regions span the 15.9 - 84.1 percentiles of the fitted profiles (approximately analogous to the 1-$\sigma$ confidence interval). The lighter shaded regions span the 2.5 - 95.5 percentiles of the fitted profiles (approximately analogous to the 2-$\sigma$ confidence interval). The best-fitting value of the equipartition parameter $\eta$ is given in the top-right corner of each plot, and the cluster name in the bottom left. These fits were made to the discrete stars. For visualisation purposes, we also show the binned velocity dispersion profiles estimated from the data as black points with error bars.}
    \label{figure:1dClassicFits}
\end{figure*}

The dispersion profile fits are shown in \autoref{figure:1dClassicFits}. To display the fits, we take the final sample from the chain and for each chain-point calculate the dispersion profile as a function of stellar mass. At each stellar mass, we then calculate the (2.5, 15.9, 50, 84.1, 97.5) percentiles. The solid yellow line shows the median profile. The dark shaded region spans the 15.9 and 84.1 percentiles (approximately analogous to the 1-sigma confidence interval) and the light shaded region spans the 2.5 to 97.5 percentiles (approximately analogous to the 2-sigma confidence region). The black points show the binned profile for visualisation. The cluster name is given in the lower-left corner and the best estimate for $\eta$ resulting from the fits is given in the upper-right corner of each panel.

\begin{figure}
    \centering
    \includegraphics[width=0.9\linewidth]{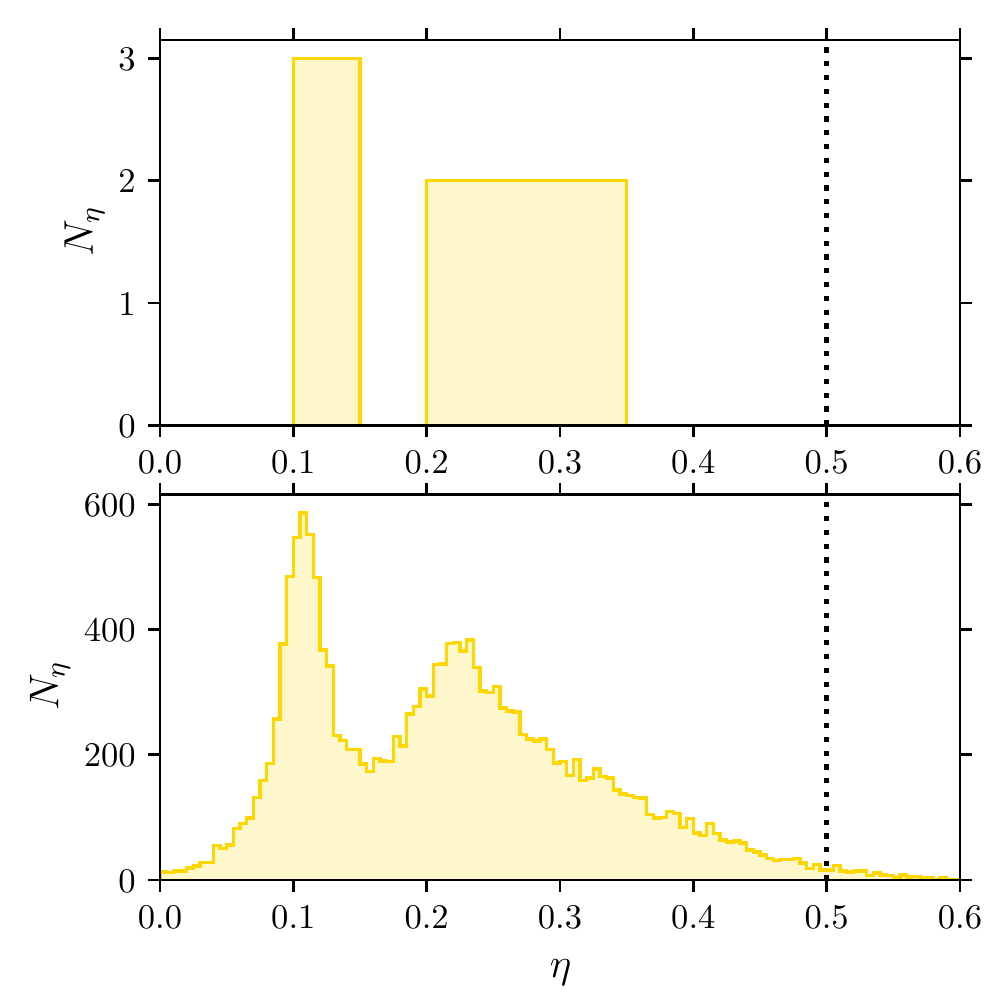}
    \caption{The distribution of equipartition parameter $\eta$ estimates from our power-law fits. Top panel: The median values estimated for each of our 9 clusters. We see two distinct peaks. No clusters reach full equipartition at $\eta = 0.5$. Bottom panel: The distribution of final chain samples (2000 points for each cluster). We still see two peaks, but with considerably more scatter. It is still exceedingly rare for the estimates to reach $\eta = 0.5$, but not completely ruled out for some clusters.}
    \label{figure:etahistogram}
\end{figure}

Histograms summarising the $\eta$ estimates for all 9 clusters are shown in \autoref{figure:etahistogram}. In the upper panel, the best estimates for $\eta$ for each of the 9 clusters are shown in bins of width 0.05. The clusters appear to separate into two groups, one spanning $\eta \sim 0.1 - 0.15$ (NGCs 5139, 6397 and 6752) and another spanning $\eta \sim 0.2 - 0.35$ (NGCs 104, 2808, 5904, 6266, 6341 and 6656); as we will discuss later in \autoref{section:correlations}, we believe this is an artefact of small number statistics and does not indicate two separate populations. The vertical dotted line marks $\eta = 0.5$, which would correspond to full equipartition. No clusters reach this value.

In the lower panel, we show the final chain samples for each cluster (2000 points for each) in bins of width 0.005 to show the full spread in $\eta$ estimates from the analysis and capture the uncertainties. Again, we see two peaks, one at $\eta \sim 0.1$ and one at $\eta \sim 0.25$. PM measurement uncertainties, and thus dispersion uncertainties and $\eta$ uncertainties will tend to increase with the heliocentric distance of the cluster. The 3 clusters in the $\eta \sim 0.1$ peak are 3 of the closer clusters in the sample, which explains the narrow peak. The peak at $\eta \sim 0.25$ has closer and more distant clusters, which contributes to the broadening of the peak. Again, the vertical dotted line marks $\eta = 0.5$, this value is not completely ruled out for the sample overall, but has very low likelihood.

\begin{deluxetable*}{c|cccc|cc|cc|cc}
\tablecaption{Summary of Catalogue Properties and Equipartition Results \label{table:summary}}
\tablehead{
    \colhead{Cluster} &
    \colhead{$N_{\star}$} &
    \colhead{$\Delta_{m}$} &
    \colhead{$\bar{R}$} &
    \colhead{$\bar{R}/R_\mathrm{half}$} &
    \colhead{$\eta$} &
    \colhead{$\meq$} &
    \colhead{Favoured Model} &
    \colhead{$R_\mathrm{disfavoured}$} &
    \colhead{$\log_{10} \Ncore$} &
    \colhead{$\log_{10} \Nhalf$} \\
    \colhead{} &
    \colhead{} &
    \colhead{(\Msun)} &
    \colhead{(arcsec)} &
    \colhead{} &
    \colhead{} &
    \colhead{(\Msun)} &
    \colhead{} &
    \colhead{} &
    \colhead{} &
    \colhead{}
}
\startdata
NGC\,104  & 16412 & 0.48 &  79.6 & 0.42 & $0.220 ^{+0.027} _{-0.024}$ & $1.37 ^{+0.16} _{-0.13}$ & Bianchini & 0.6339 & 2.23 &  0.52 \\
NGC\,2808 &  6601 & 0.16 &  65.2 & 1.36 & $0.222 ^{+0.138} _{-0.145}$ & \dots                    & \dots     & \dots  & 1.80 &  0.89 \\
NGC\,5139 & 79648 & 0.39 & 110.3 & 0.37 & $0.107 ^{+0.013} _{-0.012}$ & $2.82 ^{+0.38} _{-0.30}$ & power-law & 0.3330 & 0.48 & -0.01 \\
NGC\,5904 &  4136 & 0.26 &  57.0 & 0.54 & $0.261 ^{+0.100} _{-0.093}$ & $1.53 ^{+1.26} _{-0.51}$ & Bianchini & 0.9461 & 1.78 &  0.65 \\
NGC\,6266 &  6000 & 0.21 &  59.9 & 1.09 & $0.316 ^{+0.102} _{-0.102}$ & $1.30 ^{+0.91} _{-0.39}$ & Bianchini & 0.9471 & 2.16 &  1.08 \\
NGC\,6341 &  8200 & 0.19 &  54.1 & 0.88 & $0.311 ^{+0.089} _{-0.093}$ & $1.23 ^{+0.81} _{-0.36}$ & power-law & 0.9323 & 2.15 &  1.09 \\
NGC\,6397 &  2276 & 0.50 &  53.5 & 0.31 & $0.145 ^{+0.053} _{-0.051}$ & $1.85 ^{+1.21} _{-0.56}$ & Bianchini & 0.6490 & 5.17 &  1.51 \\
NGC\,6656 &  7118 & 0.32 &  59.6 & 0.30 & $0.256 ^{+0.051} _{-0.053}$ & $1.30 ^{+0.39} _{-0.23}$ & power-law & 0.7100 & 1.57 &  0.87 \\
NGC\,6752 &  6655 & 0.53 &  64.1 & 0.56 & $0.108 ^{+0.029} _{-0.030}$ & $2.49 ^{+1.01} _{-0.56}$ & power-law & 0.9631 & 3.22 &  1.23 \\
\enddata
\tablecomments{Summary of results. Columns: (1) cluster NGC number, (2) number of stars in final catalogue, (3) mass range between 5th and 100th percentiles in mass, (4) median radius of studied sample, (5) median radius as a fraction of the half-light radius, (6) equipartition power-law index $\eta$ estimate, (7) equipartition mass $\meq$ estimate, (8) model with higher AIC value, (9) relative likelihood of power-law model (when Bianchini model is preferred) or relative likelihood of Bianchini model (when power-law model is preferred), (10) logarithm of number of core relaxation times, (11) logarithm of number of the median relaxation times.}
\end{deluxetable*}

We provide a summary of the $\eta$ estimates and their uncertainties in  \autoref{table:summary}. We also include the number of stars in each final catalogue, the mass range spanned by the subsample used for this part of the study, and the median radius of the subsample used for this study.

\subsection{Bianchini Function Fits}
\label{section:Bianchini}

As discussed in \autoref{section:introduction}, \citet{Bianchini2016} offered a new parameterisation for velocity dispersion as a function of stellar mass (\autoref{equation:Bianchini}) and its derivative $\eta$ (\autoref{equation:BianchiniEta}). Here we wish to fit this alternative function to our samples and assess how well it describes real clusters.

Recall, this function is parameterised by an equipartition mass $\meq$ and the dispersion at the equipartition mass $\sigmaeq$. In principle, we could use these as the free parameters for our fit. In practice, these parameters are highly correlated and the locus of well-fitting points forms a curve in parameter space. This is not ideal because it makes the parameter space hard to search efficiently. Instead of $\sigmaeq$, we consider a scale dispersion $\sigmas = \sigma(\ms)$ where $\ms$ is some scale mass. $\sigmas$ will be best determined and least like to suffer high correlations with other parameters if $\ms$ lies within the mass range of the sample being fitted. The mass range spanned by each cluster is different but there is substantial overlap. The lowest upper mass limit of the sample is $\sim$0.78~\Msun and the highest lower mass limit of the sample is $\sim$0.62~\Msun. We take the middle of these values and adopt $\ms = 0.7$~\Msun for the whole sample.

Furthermore, we choose to fit in $\logmeq$ and not in $\meq$. There are two advantages here: 1) this choice naturally prevents the fit from returning negative masses, and 2) the upper mass limit is formally infinite (though we will soon restrict this with a prior) and sampling $\meq$ in log space will more efficiently sample the mass range allowed.

Now let us consider the priors on these parameters.

If all stars in a given sample are more massive than $\meq$ for that sample then all will be in full equipartition with $\eta = 0.5$, and we have no way to constrain the value of $\meq$, only to set an upper limit. This would be a problem for these fits, as any value of $\meq$ lower than the lowest mass star in the sample would be equally likely. But we know we are not in this regime, as we have already performed a power-law fit to each cluster and none returned $\eta = 0.5$. So we can place a lower limit on $\meq$ at the lowest mass in the sample for each cluster.

When all stars in a sample are less massive than $\meq$, we fare better as $\meq$ still influences the degree of equipartition in this regime. However, if $\meq$ gets infinitely large then $\eta(m) \rightarrow 0$ for all masses, and again we are unable to determine $\meq$ as all sufficiently large values will become equally likely. To mitigate this, we set an upper limit on $\meq$ at 10~\Msun.

For completeness, if $\meq$ happens to lie in the mass range of the sample under study then we would expect to be able to constrain it well and get a good fit.

So from these considerations we can set boundaries on the allowed range of $\logmeq$, but we assume a flat prior within the allowed range, that is
\begin{equation}
	P(\logmeq) = \left\{ \begin{array}{cc}
    	0 & \logmeq < \log_{10} m_\mathrm{min} \\
    	1 & \log_{10} m_\mathrm{min} \le \logmeq \le 1 \\
        0 & \logmeq > 1
    \end{array} \right.
\end{equation}

The prior on $\sigmas$ is somewhat more straightforward, and we simply insist that the velocity dispersion must be positive, but otherwise assume a flat prior, that is
\begin{equation}
	P(\sigmas) = \left\{ \begin{array}{cc}
    	1 & \qquad \sigmas \ge 0 \\
        0 & \qquad \sigmas < 0
    \end{array} \right.
\end{equation}
Although we do not set a maximum dispersion, it is worth noting that the highest dispersion would be for the lowest mass stars, but we have already set a lower mass limit, so that naturally sets an upper dispersion limit as well.

The likelihood $\mathcal{L}$ of the observed measurements given a particular model ($\logmeq$, $\sigmas$) is once again given by \autoref{equation:likelihood}, where in this case we use $\sigma_i = \sigma(m_i | \logmeq, \sigmas)$ for star $i$. Then the posterior probability $\mathcal{P}$ of a model ($\logmeq$, $\sigmas$) given the observed data is given by
\begin{equation}
    \mathcal{P} = \mathcal{L} P(\logmeq) P(\sigmas) .
\end{equation}

Once again, we use MCMC via \textsc{emcee} to efficiently sample the parameter space. We run the chains with 100 walkers for 1000 steps and take 2000 points (taken from 20 steps sampled in 10-step intervals for the end of the chain) to define the sample.

\begin{figure}
    \centering
    \includegraphics[width=\linewidth]{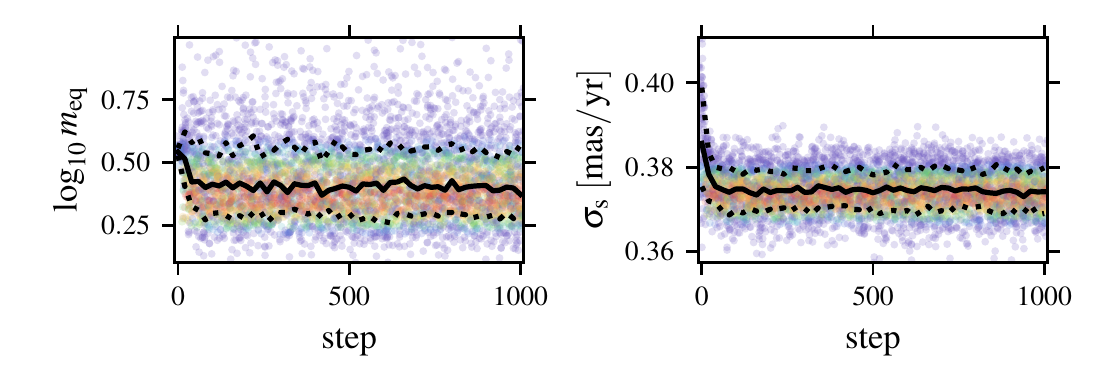}
    \includegraphics[width=0.9\linewidth]{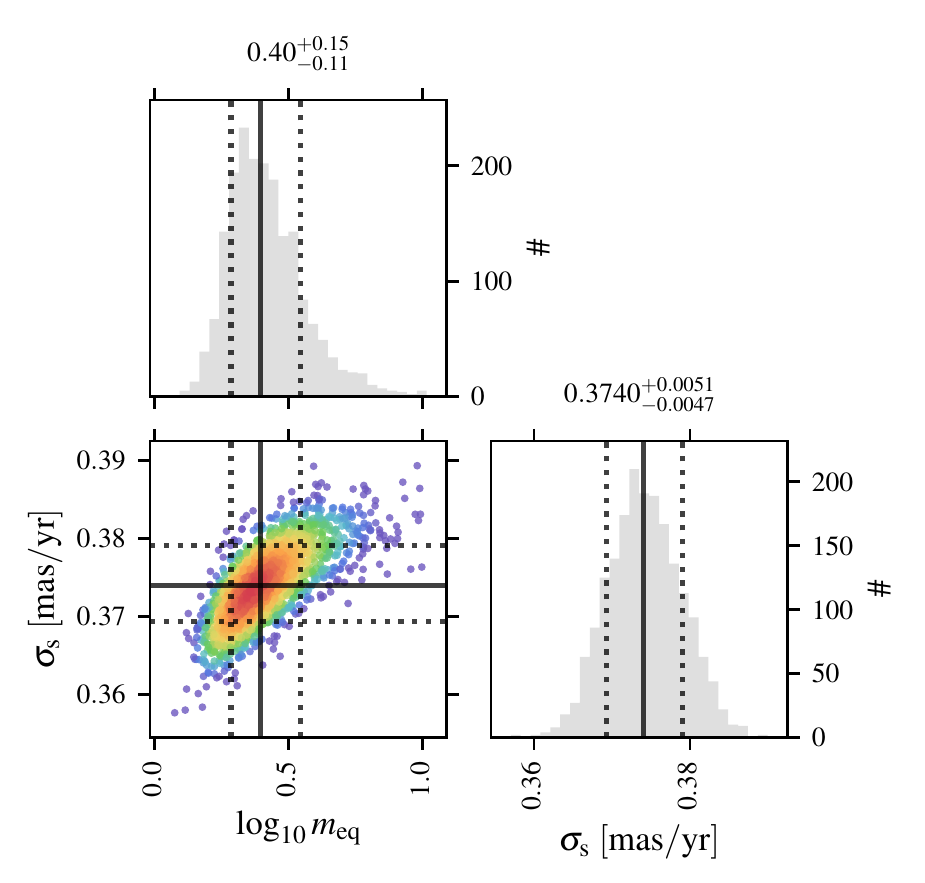}
    \caption{MCMC chain results for the 1d \citet{Bianchini2016} fit to NGC\,6752. Top panels: progression of the two free parameters $\logmeq$ (left) and $\sigmas$ (right) for the whole run. Both panels are flat showing that the chain has converged well. Bottom panels: distribution of walkers at the end of the run. The histograms show the distribution of the individual $\logmeq$ and $\sigmas$ parameters and are approximately Gaussian; the scatter plot shows their correlation in phase-space.}
    \label{figure:1dBianchiniChain}
\end{figure}

All fits converged. As an example, \autoref{figure:1dBianchiniChain} shows the location of the walkers as the chain progressed (top panels) and the distribution of points in phase space for the final sample. For 8 of the 9 clusters, the final sample does not hit the limits of the priors that we set and the plots look very similar to this.

\begin{figure}
    \centering
    \includegraphics[width=0.9\linewidth]{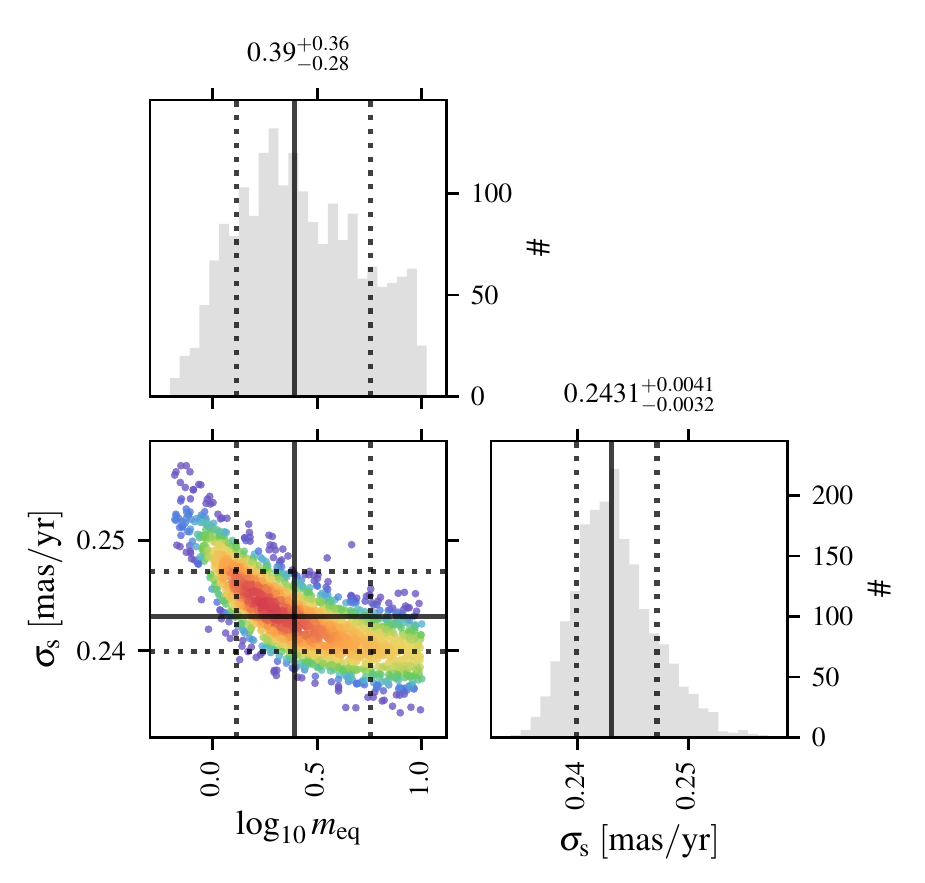}
    \caption{The same as the bottom panels of \autoref{figure:1dBianchiniChain} for NGC\,2808. This fit was unsuccessful as it hit the prior boundary at $\logmeq = 1$.}
    \label{figure:1dBianchiniChain2808}
\end{figure}

\autoref{figure:1dBianchiniChain2808} shows the corresponding final parameter distribution plots for NGC\,2808. The region of highest likelihood (coloured red-orange) lies within the allowed parameter space, but the high-mass tail of the distribution does hit the upper edge of the prior at 10~\Msun. Any $\meq$ higher than 10~\Msun will have the same likelihood as found at the prior boundary. This also means that the histogram of $\logmeq$ would have an infinite high-mass tail had we not set the prior. We omit this cluster from this part of the analysis.

\begin{figure*}
    \centering
    \includegraphics[width=0.3\linewidth]{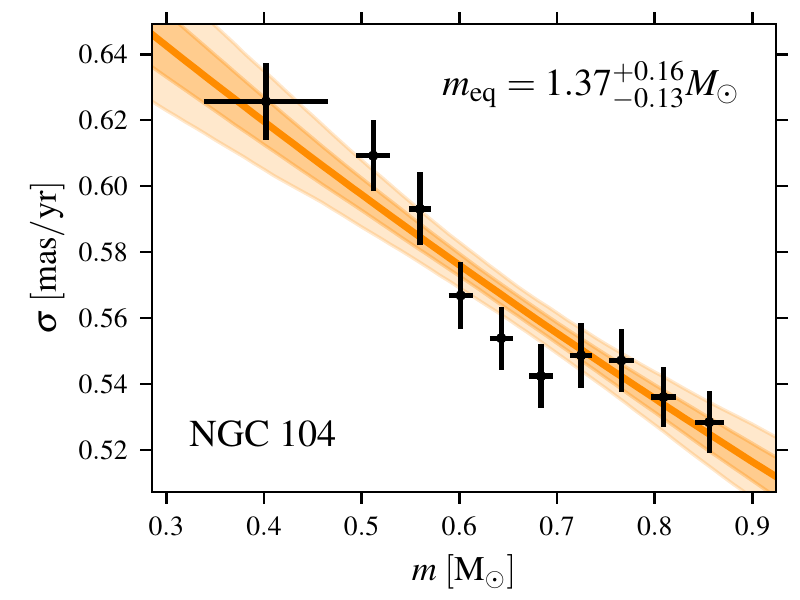}
    \includegraphics[width=0.3\linewidth]{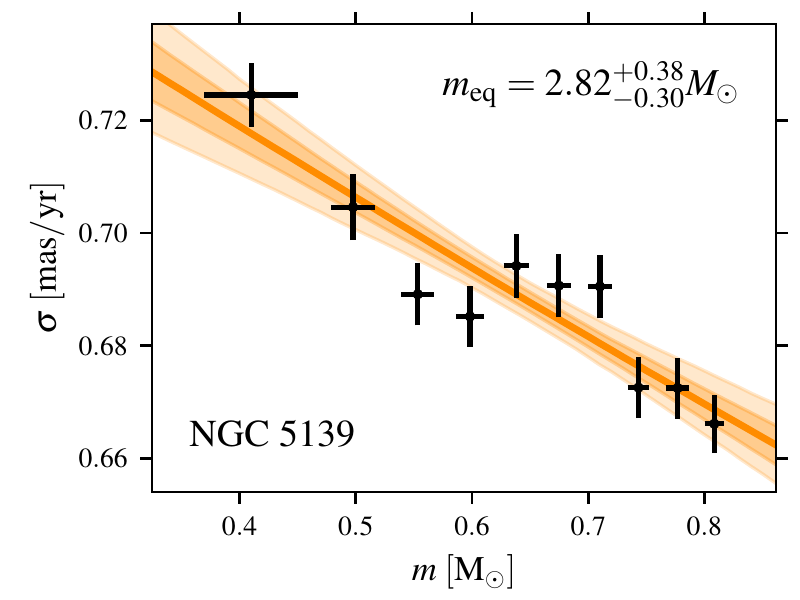}
    \includegraphics[width=0.3\linewidth]{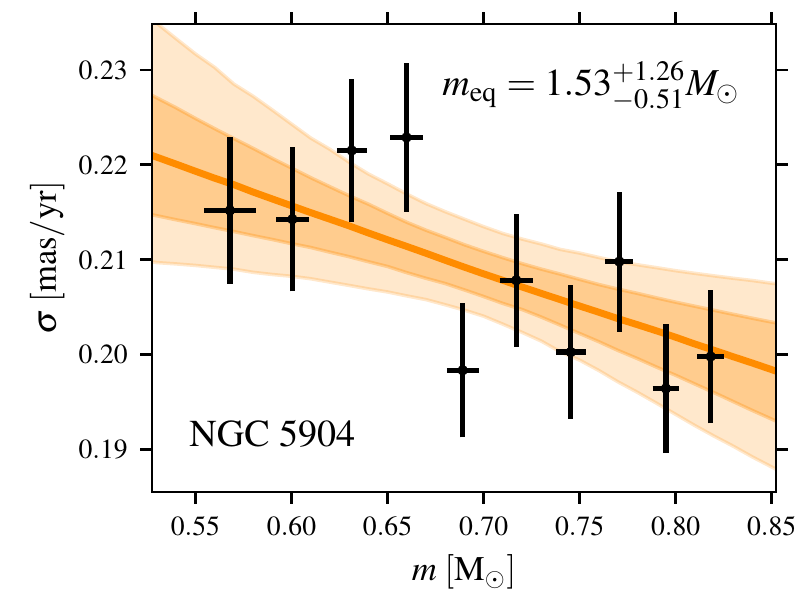}
    \includegraphics[width=0.3\linewidth]{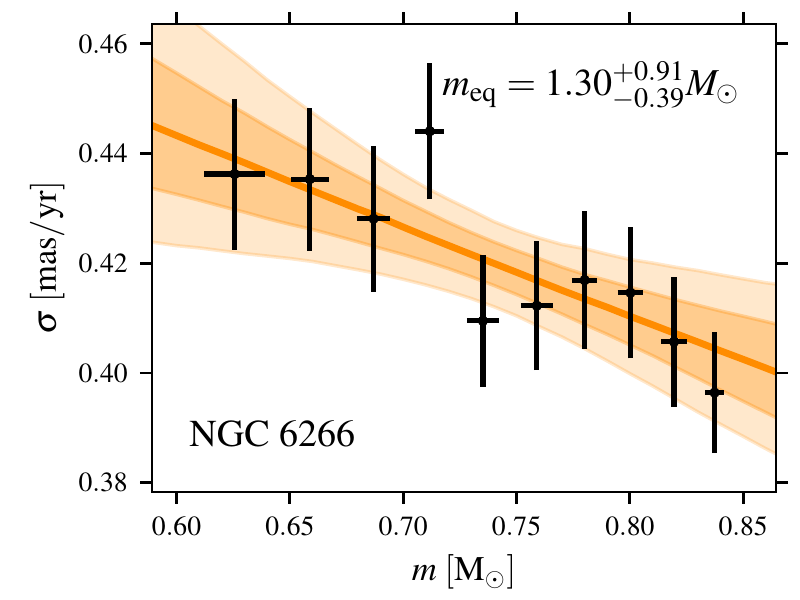}
    \includegraphics[width=0.3\linewidth]{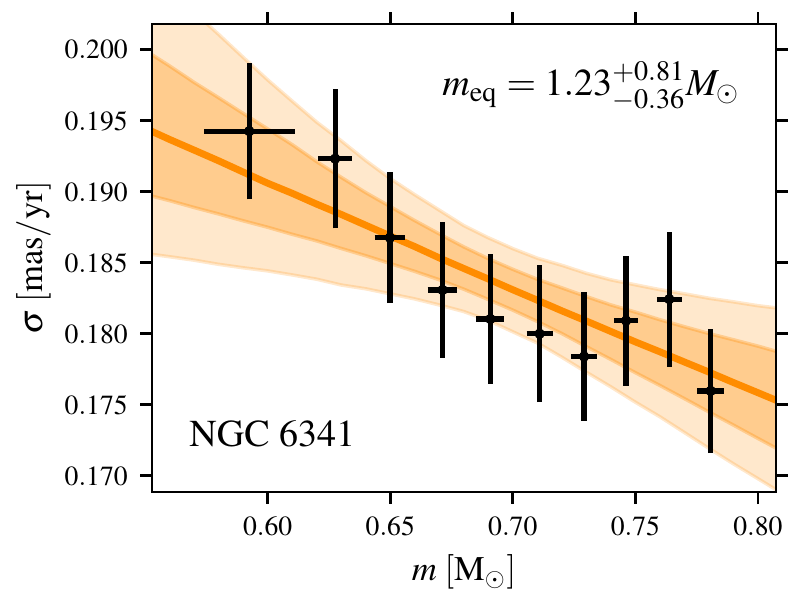}
    \includegraphics[width=0.3\linewidth]{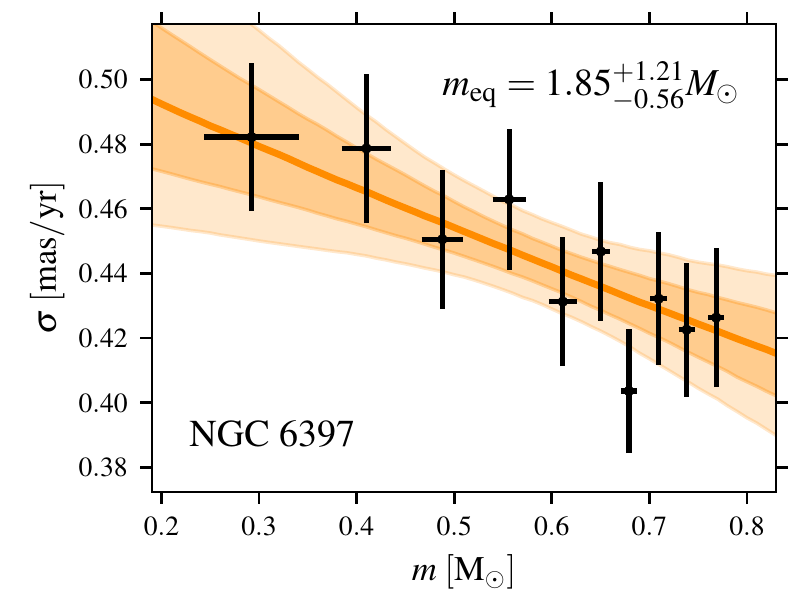}
    \includegraphics[width=0.3\linewidth]{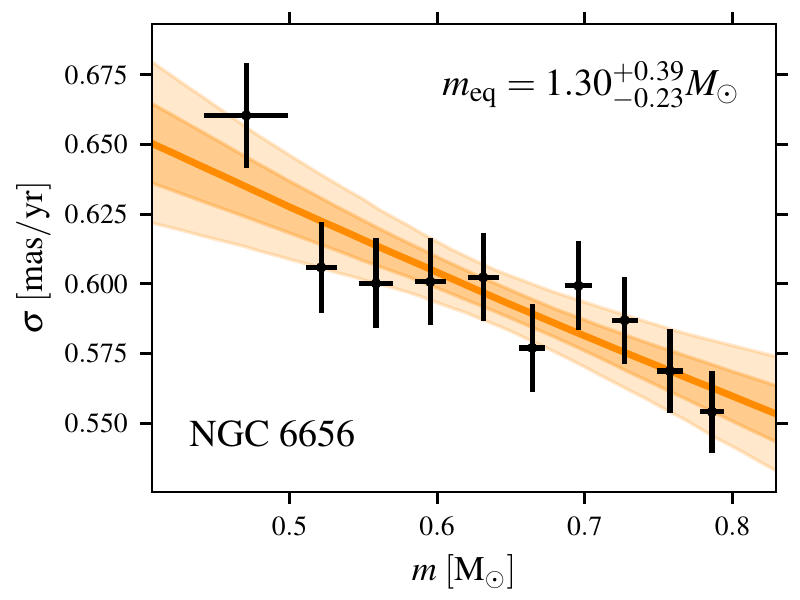}
    \includegraphics[width=0.3\linewidth]{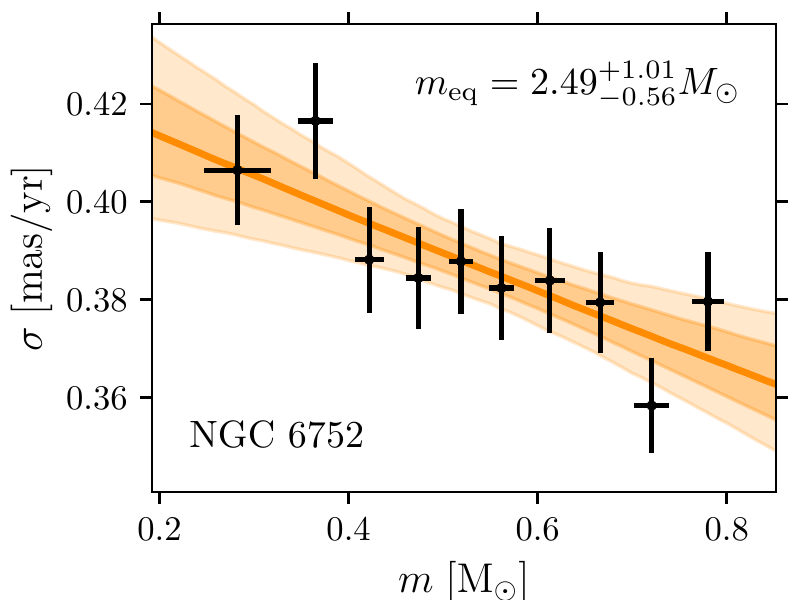}
    \caption{\citet{Bianchini2016} fits to the velocity dispersion as a function of stellar mass for 8 of the 9 clusters in our sample (the fit for NGC\,2808 was unsuccessful). The solid orange lines show the median of the fitted profiles. The darker shaded regions span the 15.9 - 84.1 percentiles of the fitted profiles (approximately analogous to the 1-$\sigma$ confidence interval). The lighter shaded regions span the 2.5 - 95.5 percentiles of the fitted profiles (approximately analogous to the 2-$\sigma$ confidence interval). The best-fitting value of the equipartition mass $\meq$ is given in the top-right corner of each plot, and the cluster name in the bottom left. These fits were made to the discrete stars. For visualisation purposes, we also show the binned velocity dispersion profiles estimated from the data as black points with error bars.}
    \label{figure:1dBianchiniFits}
\end{figure*}

In \autoref{figure:1dBianchiniFits}, we show the resulting fits for the 8 successful clusters. Once again, we take the final sample from the MCMC chain for each cluster and calculate the dispersion profile for each parameter position in the sample. Then we calculate the (2.5, 15.9, 50, 84.1, 97.5) percentiles at each point along the dispersion profile. The orange lines show the median of the dispersion profiles. The dark shaded areas indicate the 15.9--84.1 percentile interval (approximately analogous to the 1-$\sigma$ confidence interval) and the light shaded areas indicate the 2.5--97.5 percentile interval (approximately analogous to the 2-$\sigma$ confidence interval). The black points with error bars are the binned dispersion profile we calculated earlier -- again, these binned points were generated for visualisation purposes. The fits were performed discretely using the individual stars. The cluster name is given in the lower-left corner and the best estimate for $\meq$ in the upper right.

As the clusters are all $\sim$11--13~Gyr, the most massive stars in the cluster samples are typically 0.8--0.9~\Msun. Note that all of the equipartition mass $\meq$ estimates are more massive than this, indicating that the clusters are not in full equipartition, but only partial equipartition. This conclusion is consistent with the classic power-law fit above.

\begin{figure}
    \centering
    \includegraphics[width=0.9\linewidth]{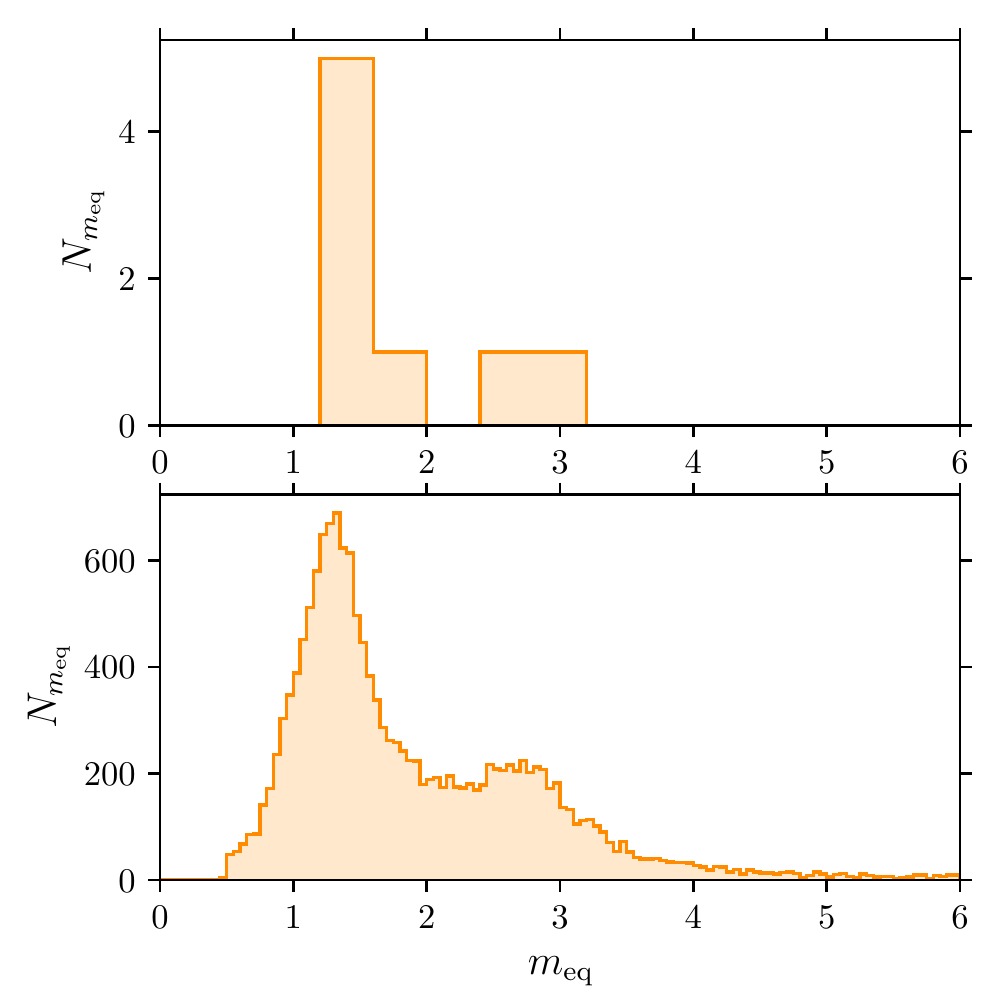}
    \caption{The distribution of equipartition parameter $\meq$ estimates from our \citet{Bianchini2016} fits. Top panel: The median values estimated for each of the 8 clusters. We see two distinct peaks, as we did for the $\eta$ distribution. Bottom panel: The distribution of final chain samples (2000 points for each cluster). Again, there are hints of two peaks here.}
    \label{figure:meqhistogram}
\end{figure}

Histograms summarising the $\eta$ estimates for all 8 clusters are shown in \autoref{figure:meqhistogram}. As before, the upper panel shows the best estimates for $\meq$ for each of the 8 clusters, shown in bins of width 0.4. Again, the clusters appear to separate into two groups, one spanning $\meq \sim 1.2 - 2$ (NGCs 104, 5904, 6266, 6341, 6397 and 6656) and another spanning $\meq \sim 2.4 - 2.9$ (NGCs 5139 and 6752), but this is likely an artefact of small number statistics and does not indicate real groupings (see \autoref{section:correlations}). In the lower panel, we show the final chain samples for each cluster (2000 points for each) in bins of width 0.05 to show the full spread in $\meq$ estimates from the analysis and capture the uncertainties. Again, we see two peaks, one at $\meq \sim 1.4$ and one at $\eta \sim 2.8$.

We provide a summary of the $\meq$ estimates and their uncertainties in  \autoref{table:summary}.

\subsection{Comparison of 1D Fits}
\label{section:1dcomparison}

The next question we wish to ask is how these two different fitting functions compare. Were we to plot the dispersion profile fits on top of one another they would be largely indistinguishable, at least visually, as both have managed to reproduce the dispersion as a function of stellar mass well.

\begin{figure*}
    \centering
    \includegraphics[width=0.3\linewidth]{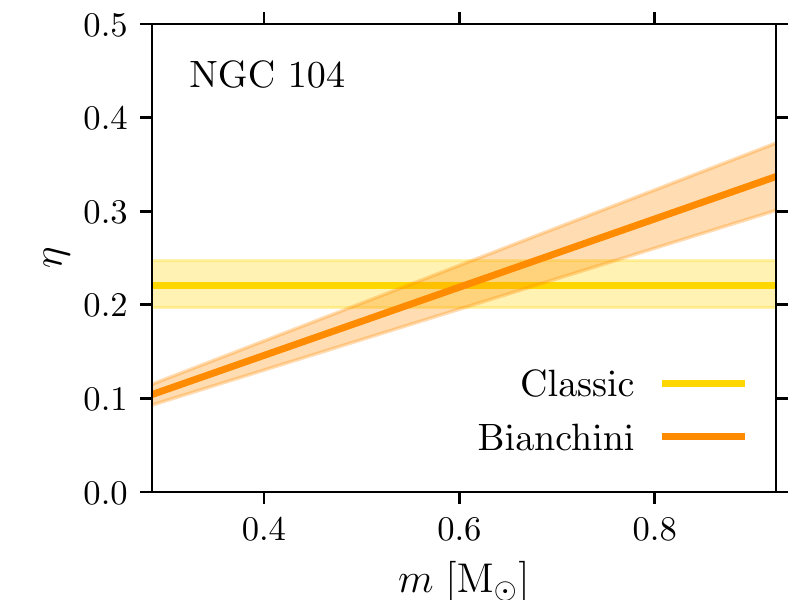}
    \includegraphics[width=0.3\linewidth]{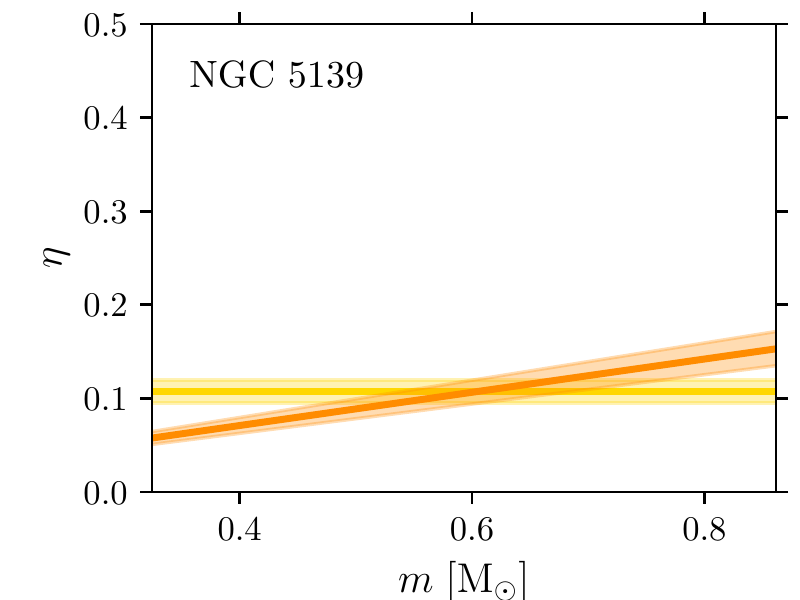}
    \includegraphics[width=0.3\linewidth]{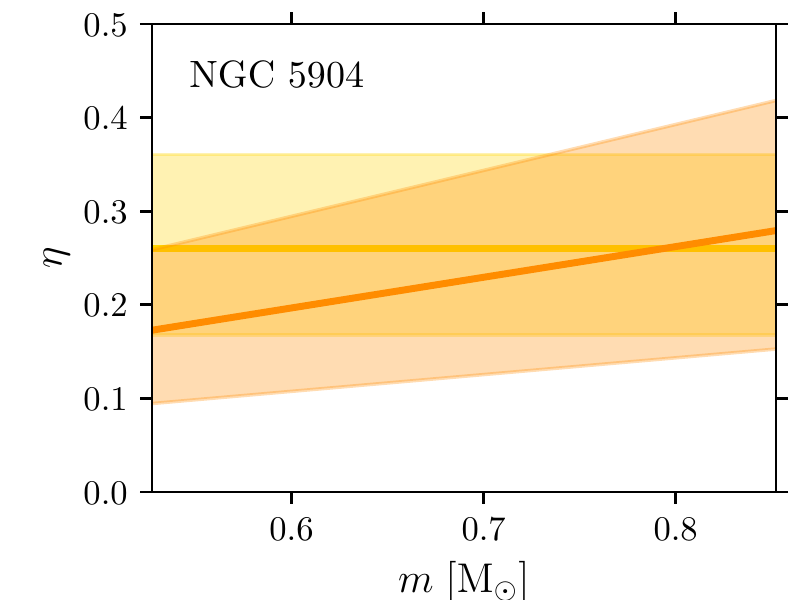}
    \includegraphics[width=0.3\linewidth]{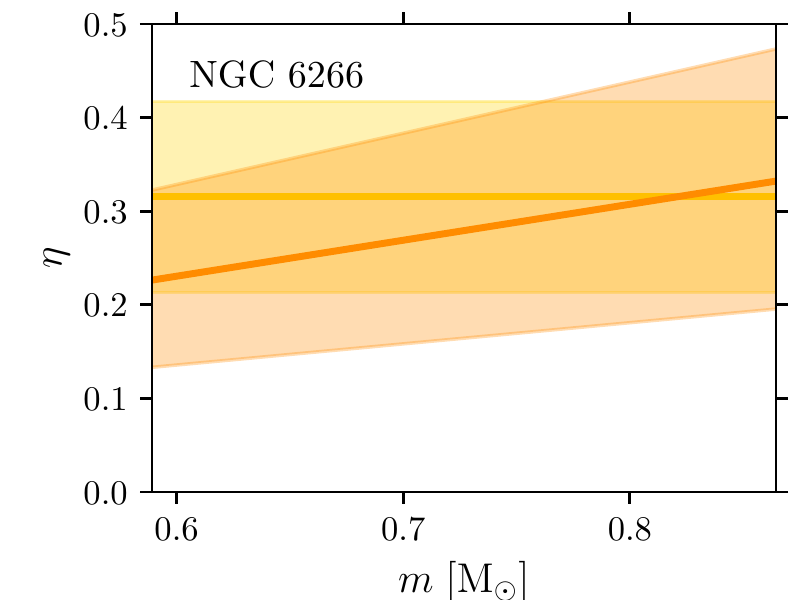}
    \includegraphics[width=0.3\linewidth]{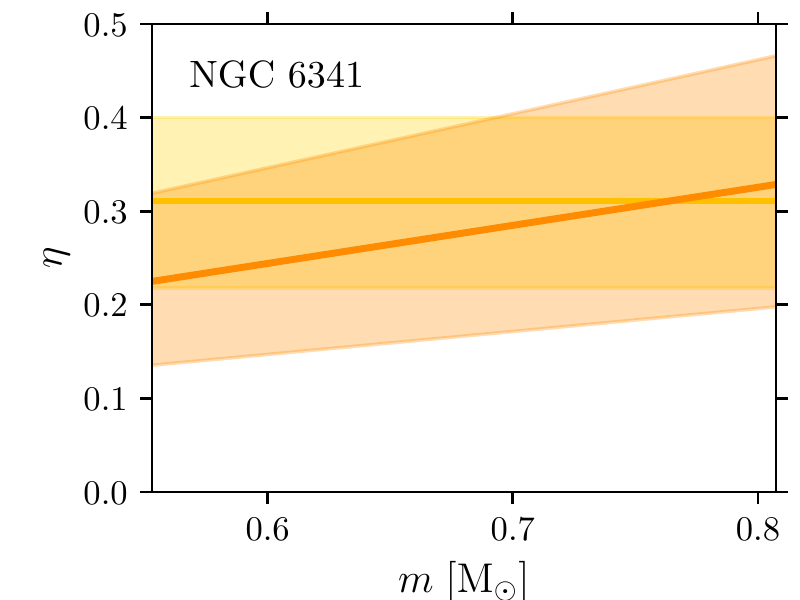}
    \includegraphics[width=0.3\linewidth]{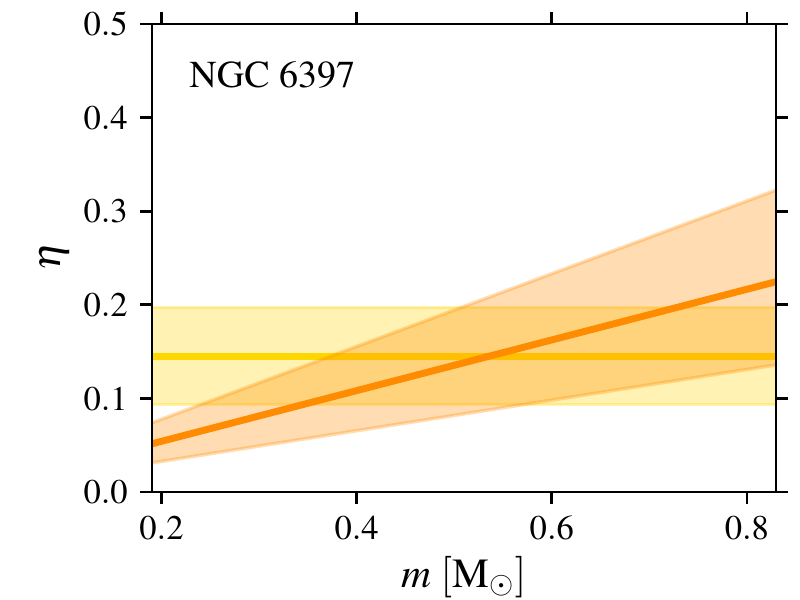}
    \includegraphics[width=0.3\linewidth]{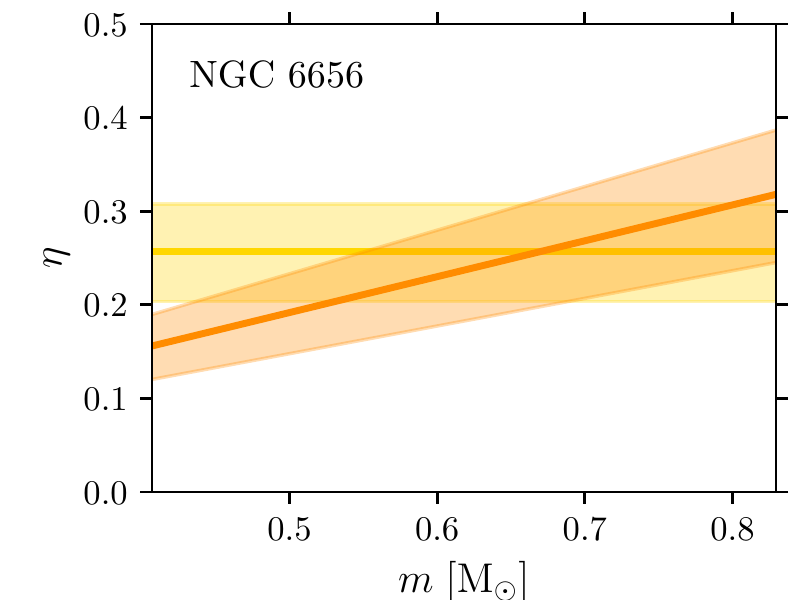}
    \includegraphics[width=0.3\linewidth]{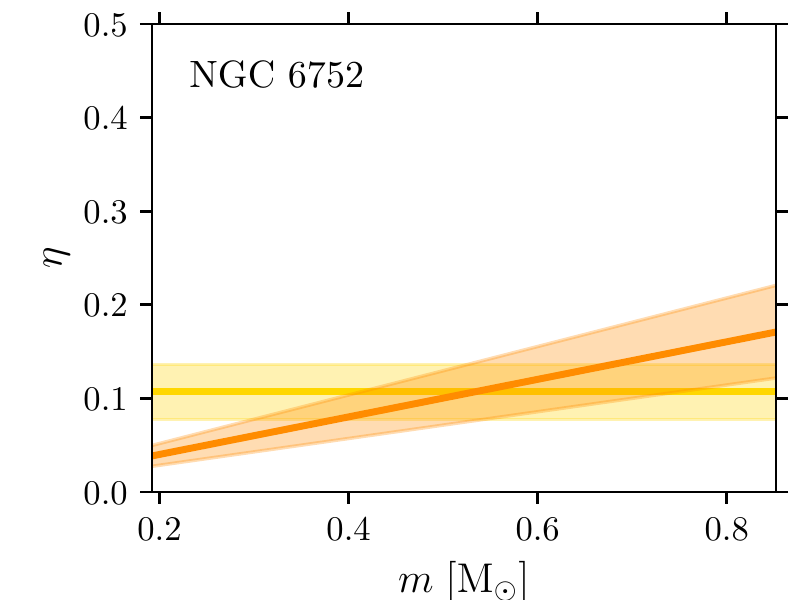}
    \caption{Comparison of equipartition $\eta$ as a function of stellar mass for the 8 clusters for which both fits could be performed. The yellow lines and shaded regions show the median $\eta$ from the power-law fits and the 15.9-84.1 percentile intervals. These lines are constant by definition. The orange lines and shaded regions shows the median $\eta$ from the \citet{Bianchini2016} fits and the 15.9-84.1 percentile intervals. These are straight lines with slope $\frac{1}{2\meq}$ for $\meq$ the \citet{Bianchini2016} equipartition parameter.}
    \label{figure:EtaMass}
\end{figure*}

Instead, let us consider the degree of equipartition $\eta$ as a function of stellar mass for the two fits. For the power-law fit, this is flat by definition. For the Bianchini fit, we can use \autoref{equation:BianchiniEta}. The resulting profiles are shown in \autoref{figure:EtaMass}. The solid lines show the median $\eta$ values and the shaded regions show the 15.9--84.1 percentile interval (analogous to the 1-$\sigma$ confidence interval)\footnote{We do not show the 2.5--97.5 percentile intervals on these plots as they become too cluttered.}. Yellow shows the classic power-law fit and orange shows the Bianchini function fit. Each cluster is shown on the same $\eta$ scale of 0--0.5.

We see here again that no cluster reaches full equipartition at $\eta = 0.5$. Pleasingly, we also see that the $\eta$ profiles for the two different fits are broadly consistent, even if their details differ. They tend to agree more at the high-mass end than the low-mass end as the uncertainties are smaller for higher masses.

But is one model preferred over the other? To answer this question we compute the Akaike Information Criterion \citep[AIC,][]{Akaike1974}\footnote{In this case, this is equivalent to computing the Bayesian Information Criterion \citep[BIC,][]{Schwarz1978}. The two tests penalise number of free parameters in different ways, but we have two free parameters in both models so the penalty term is the same for both models.}
\begin{equation}
    \mathrm{AIC} = 2 N - 2 \ln \left( \mathcal{L}_\mathrm{max} \right)
\end{equation}
for both models for each cluster, where $N$ is the number of free parameters in the model and $\mathcal{L}_\mathrm{max}$ the maximum likelihood for the model fit. The model with the lowest AIC value is the preferred model.

\autoref{table:summary} lists the preferred model for each of the 8 clusters for which we could perform this analysis. Four clusters favour the classic power-law fit, and four clusters favour the Bianchini function fit.

We can also calculate the relative likelihood of each model by calculating $R = \exp \left( \left( \mathrm{AIC}_\mathrm{min} - \mathrm{AIC}_i \right) / 2 \right)$ where $\mathrm{AIC}_\mathrm{min}$ is the minimum of the two model AIC values, and $\mathrm{AIC}_i$ is the AIC for model $i$. This will give $R=1$ for the favoured model, and $R<1$ for the less-preferred model; we provide the latter in \autoref{table:summary}. A smaller $R$ value indicates a more significant difference.

Four clusters have $R >$~0.9, indicating that the differences are negligible. Three clusters have $R \sim 0.65$ and one has $R \sim 0.35$ -- these values indicate more meaningful differences but are not at a level that would allow us to rule out either model for any cluster. More data is needed to determine which functional form provides the better fit.

\section{Equipartition Profiles}
\label{section:etaprofiles}

So far we have neglected the radial distance of each star within the cluster, save to make an initial cut to select the stars between the 25th and 75th percentiles in radius. So the results above could have a subtle radial effect folded into them as well. Moreover, we expect equipartition and related effects to decrease with increasing distance from the cluster centre \citep[e.g.][]{Bianchini2018}.

Here, we instead bin the stars in radius and then study how the dispersion changes as a function of stellar mass in each radial bin. This is more challenging, as we will have fewer stars per bin than we did our initial analysis, which is why we didn't do this from the start.

We compute the number of bins needed to have 2500 PM measurements per bin (remember there are 2 PM measurements per star). Where this is fewer than 3, we use 3 bins, because to have a meaningful profile we need at least 3 points. Where it is more than 10, we use 10 bins, to avoid the signal being overwhelmed by bin-to-bin scatter.

Now we have a choice whether to use the classic power-law fit or the Bianchini function fit to estimate the degree of equipartition in each bin. We saw in \autoref{section:1dcomparison} that neither model is favoured over the other for the whole sample, so we elect to use the classic power-law fit as it is intuitively easier to interpret.

So, in each bin, we proceed as for \autoref{section:classic} above and use MCMC to estimate the velocity dispersion of a 1~\Msun star $\sigma_1$ and the degree of equipartition $\eta$ in each bin. We use the same priors as before to restrict the ranges of $\eta$ and $\sigma_1$. There are no further priors that govern the behaviour of adjacent bins so, for each cluster, the fits in each bin are independent of the fits in all other bins.

We run the MCMC chains with 100 walkers for 5000 runs. The final sample is again composed of 2000 points taken from the end of the chain, this time from 20 steps taken in 50-step intervals.

\begin{figure*}
    \centering
    \includegraphics[width=0.49\linewidth]{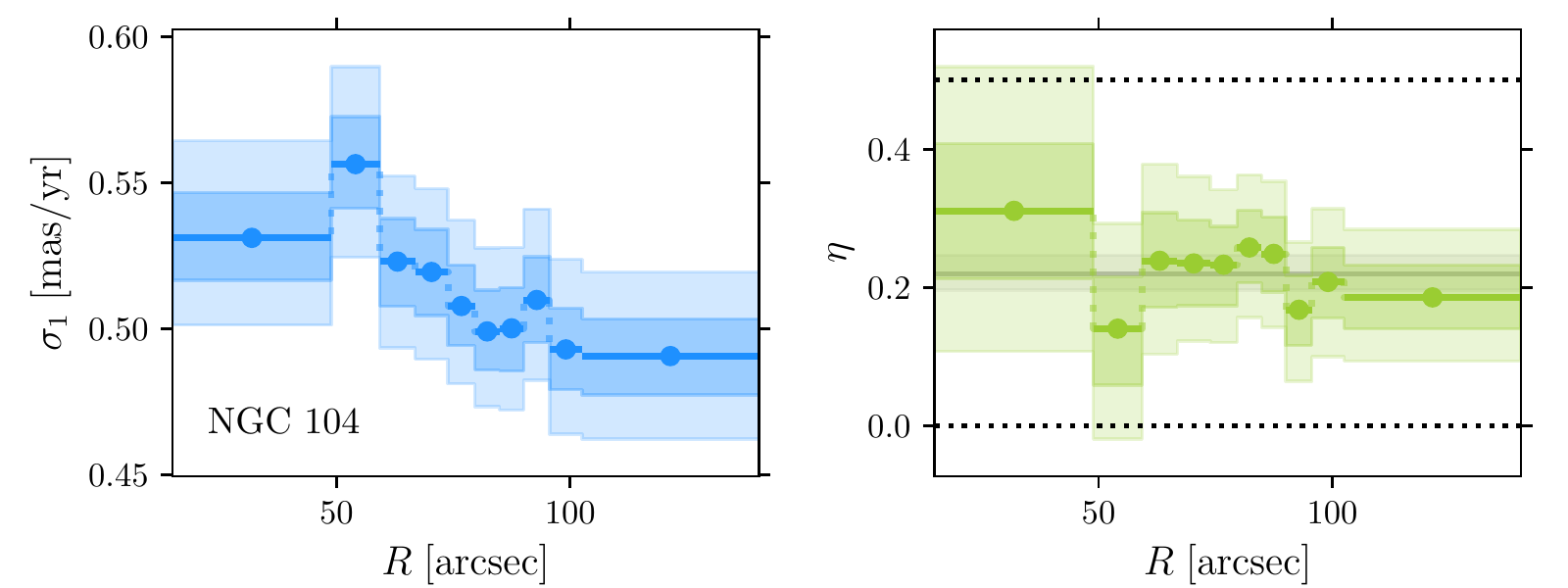}
    \includegraphics[width=0.49\linewidth]{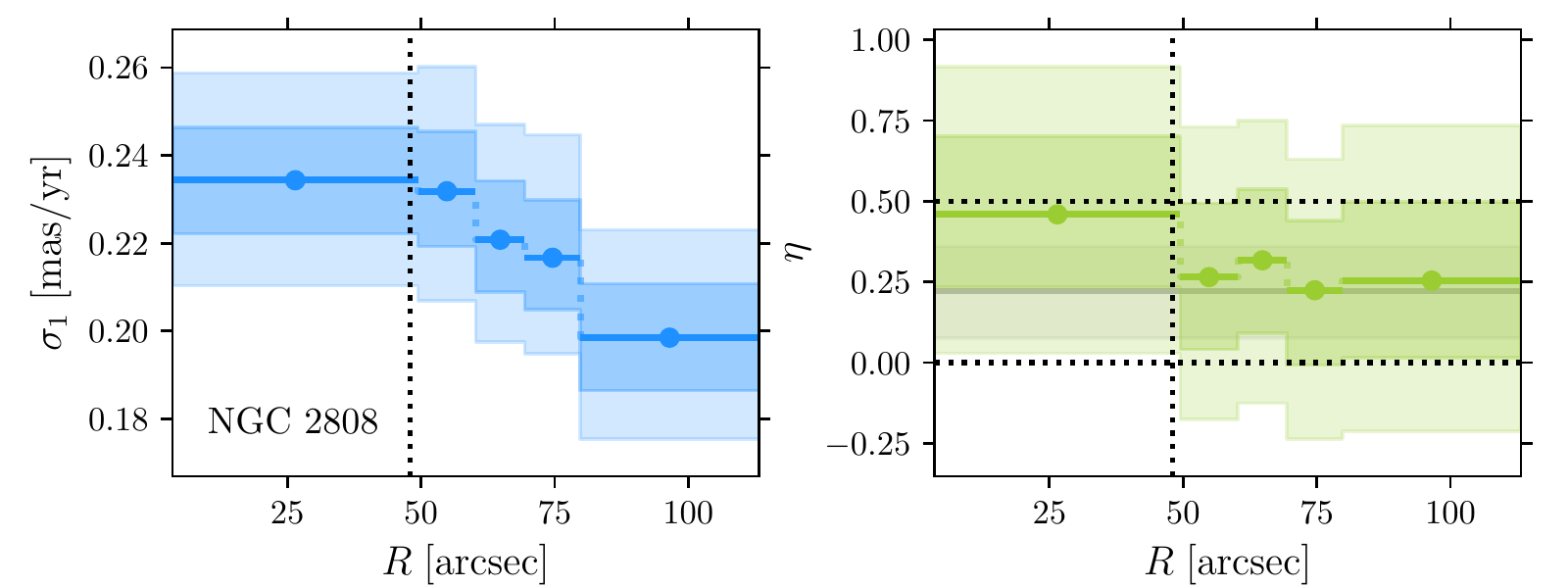}
    \includegraphics[width=0.49\linewidth]{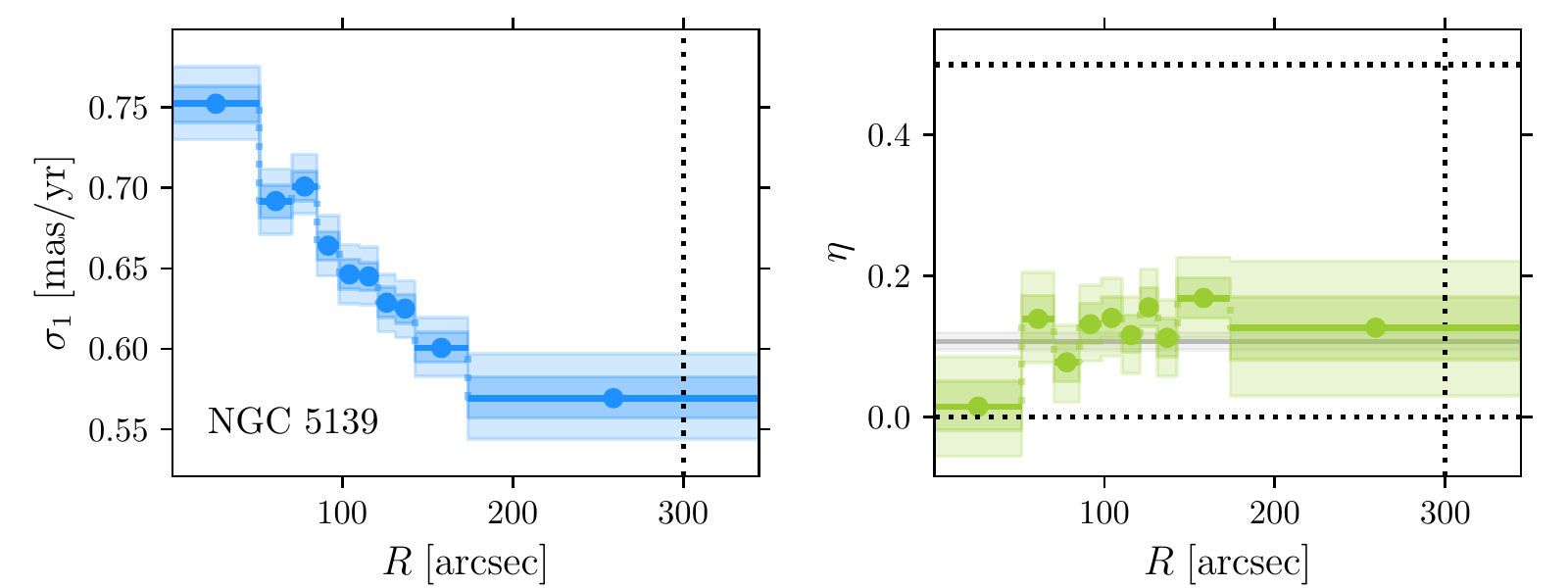}
    \includegraphics[width=0.49\linewidth]{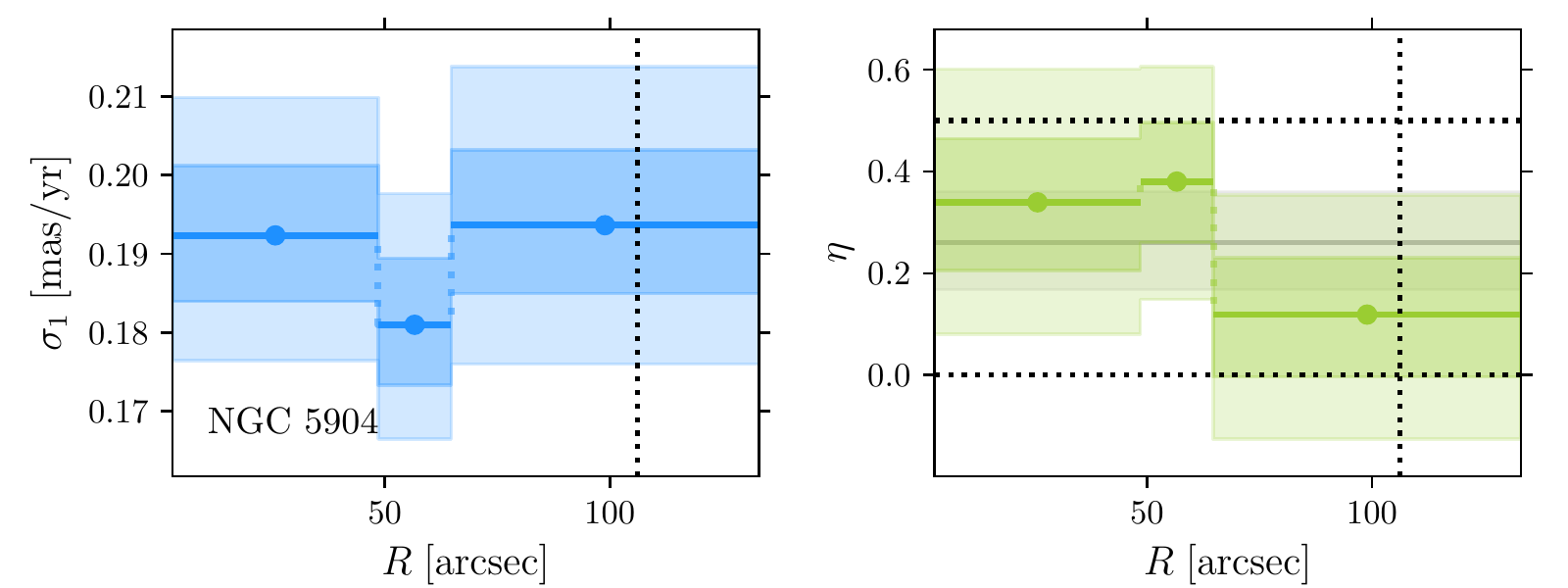}
    \includegraphics[width=0.49\linewidth]{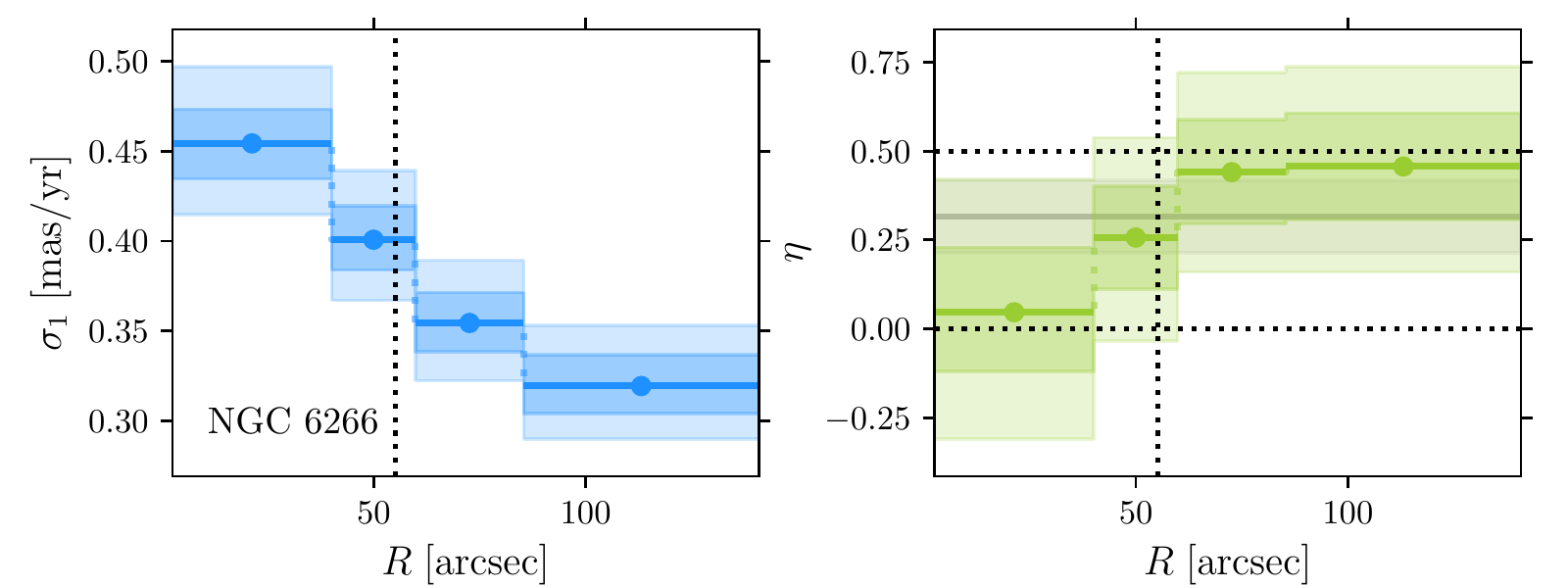}
    \includegraphics[width=0.49\linewidth]{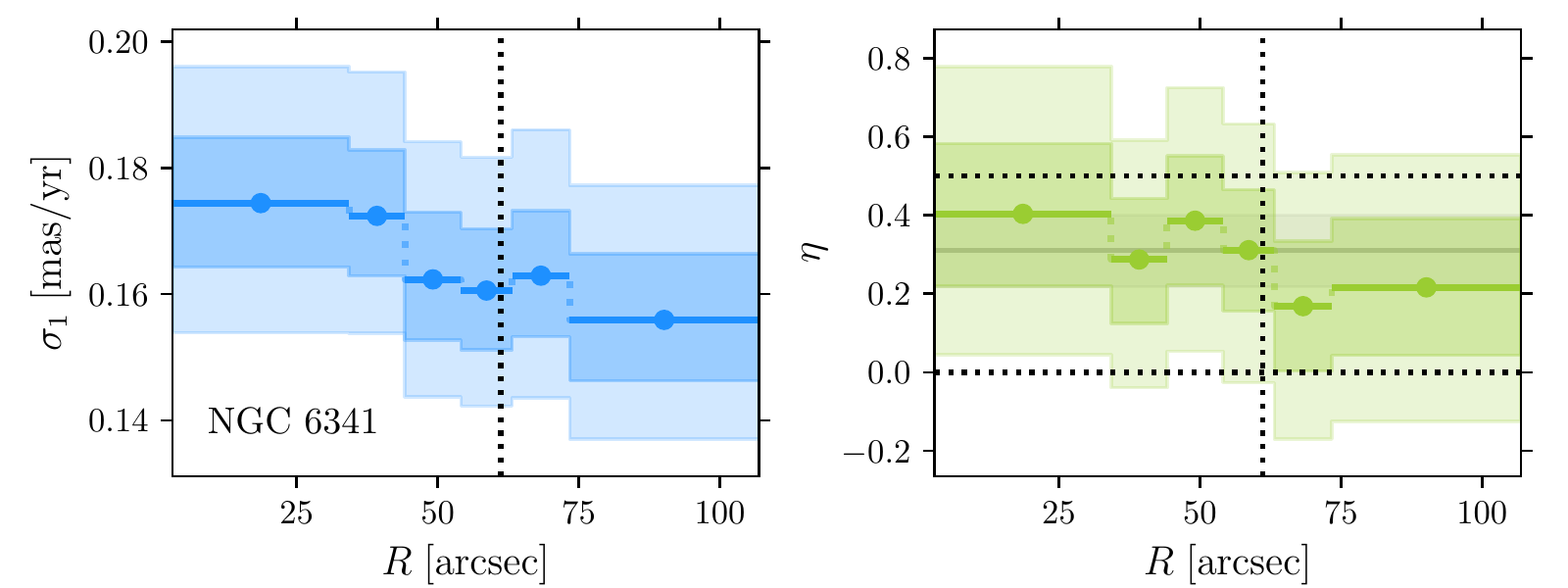}
    \includegraphics[width=0.49\linewidth]{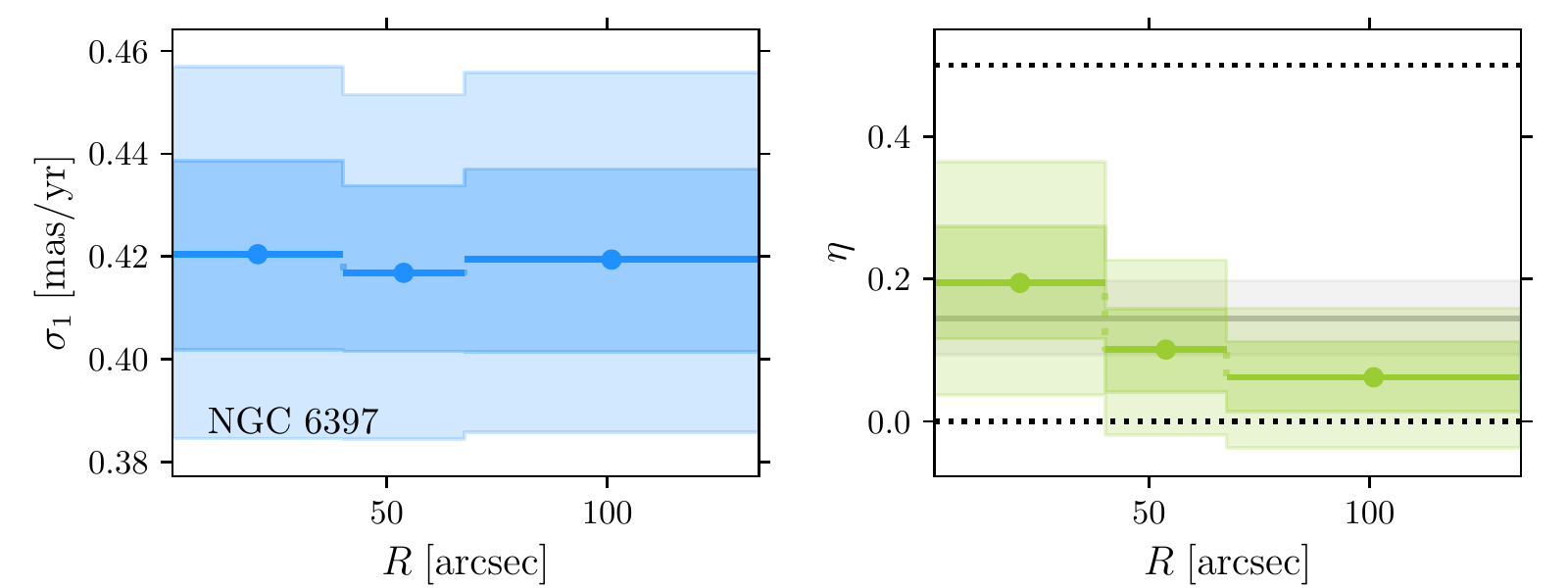}
    \includegraphics[width=0.49\linewidth]{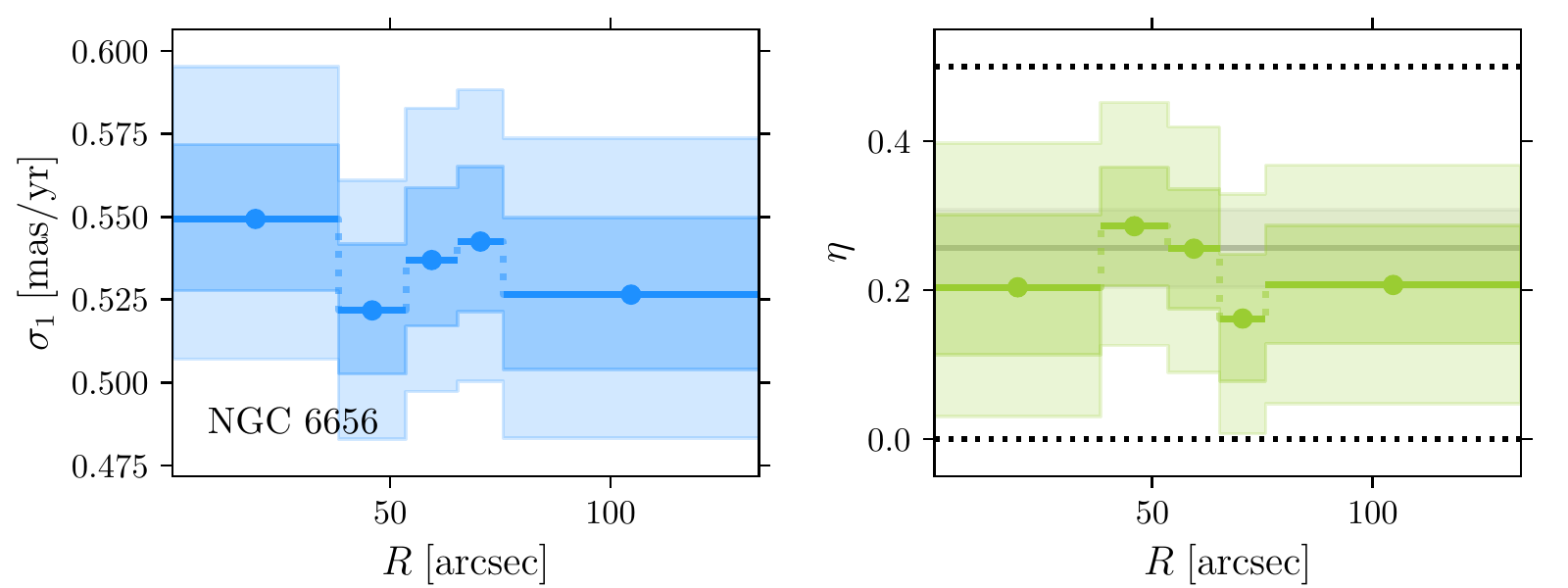}
    \includegraphics[width=0.49\linewidth]{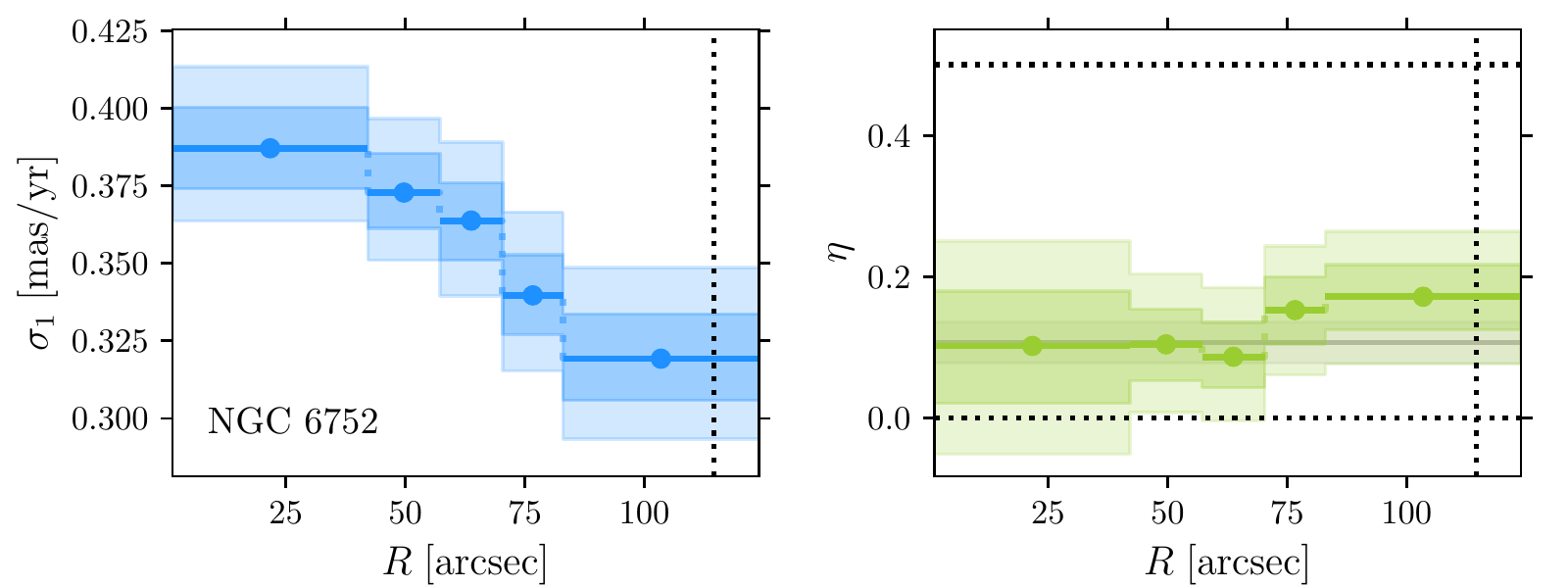}
    \hspace{0.49\linewidth}
    \caption{Equipartition radial profiles. Left panels: Velocity dispersion of a 1~\Msun star $\sigma_1$ as a function of distance from the cluster centre (shown in blue). Right panels: Equipartition parameter $\eta$ as a function of distance from the cluster centre (shown in green). In both panels, the solid line shows the median of the fit in each radial bin, the darker shaded region spans the 15.9-84.1 percentile interval (approximately analogous to the 1-$\sigma$ confidence region), and the lighter shaded region spans the 2.5-95.5 percentile interval (approximately analogous to the 2-$\sigma$ confidence region). The dots mark the bin centres. Horizontal dotted lines mark $\eta = 0$ (no equipartition) and $\eta = 0.5$ (full equipartition). Where they fall within the fields of view, vertical dotted lines mark the half-light radii $\Rhalf$. In general, the velocity dispersion profiles are highest at the centre and fall off with radius, as expected. The equipartition profiles are broadly flat with radius.}
    \label{figure:EtaRadius}
\end{figure*}

\autoref{figure:EtaRadius} shows the results for each cluster. In the left-hand panels (in blue), we show how $\sigma_1$ changes as a function of radius. In the right-hand panels (in green), we show how $\eta$ changes as a function of radius. In both panels, the solid lines show the median values with the coloured point marking the bin centre and the length of the line showing the bin width. The dark shaded regions show the 15.9 -- 84.1 percentile interval (approximately analogous to the 1-$\sigma$ confidence interval), and the light shaded regions show the 2.5 -- 97.5 percentile interval (approximately analogous to the 2-$\sigma$ confidence interval). In the right-hand panels, the black dotted lines mark $\eta = 0$ (no equipartition) and $\eta = 0.5$ (full equipartition) when these value lies within the plot limits. The grey solid lines and grey shaded regions show the best fitting values of $\eta$ with their uncertainties from the fits in \autoref{section:classic} for comparison.

We would generally expect the dispersion profiles to decrease with increasing radius, which is broadly the behaviour we observe here, albeit with significant noise for some clusters. Cluster dispersion profiles have been well studied in the past, so this is not a particularly interesting feature of this work. We include these only to show that they are consistent with our expectations and demonstrate that the fits are reasonable.

The interesting part of these fits are the $\eta$ profiles. From a visual inspection, there are hints of some trends in $\eta$ with radius in some panels, but within the uncertainties all are consistent with being flat as a function of radius. It would be nice to frame this in a more statistical basis.

To that end, we fit to the $\eta$ profiles a zeroth-order and first-order polynomial, accounting for the uncertainties. Similarly to \autoref{section:1dcomparison} we also calculate the AIC for the two fits. Two clusters -- NGC\,5139 and NGC\,6266 -- prefer the first-order polynomial, albeit at low significance. That is, the models show a slight preference for $\eta$ that changes with radius (in both cases, in fact, $\eta$ that increases with radius), but a flat profile cannot be ruled out. For the remaining 7 clusters, we find that the zeroth-order polynomial fit is preferred, albeit again at low significance. That is, that the model prefers the fit whereby $\eta$ does not change with radius, although a changing profile cannot be ruled out.

Perhaps the strongest conclusion we can draw here is that such an analysis is pushing the limits of the data that we have at present, but does offer hope for the future as datasets improve.

An important caveat to note is that these are all central fields. The exact extent of the coverage depends on the cluster's intrinsic size and heliocentric distance, and the location of the observed fields, but no fields cover the full extent of any cluster. So whatever conclusions we draw here only apply to these central regions; the $\eta$ profiles could behave differently in regions beyond these fields.

\section{Discussion}
\label{section:discussion}

\subsection{Correlations with Cluster Properties}
\label{section:correlations}

Now that we have estimated $\eta$ and $\meq$ for each of the clusters, we can search for correlations between $\eta$ or $\meq$ and other cluster properties. Though let us first consider whether these quantities are correlated themselves. Given that they are probing the same fundamental process (and based on the consistency shown in \autoref{figure:EtaMass}), we would certainly expect them to show some degree of correlation.

\begin{figure}
    \centering
    \includegraphics[width=0.9\linewidth]{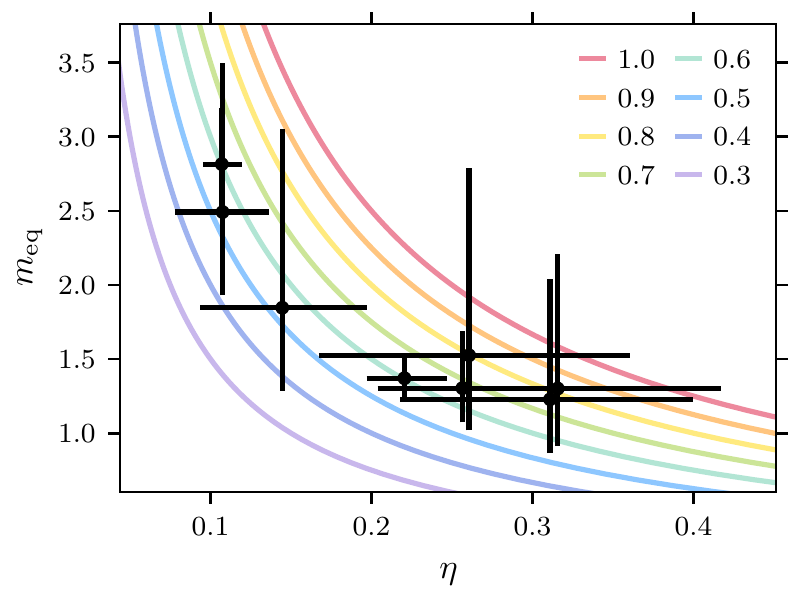}
    \caption{Estimates of equipartition $\eta$ versus estimates of equipartition mass $\meq$ for the two global fits. Black points with error bars show the estimates. The lines show how $\eta$ and $\meq$ are related for different stellar masses from \autoref{equation:BianchiniEta}, coloured by mass according to the legend. Proximity to a given line roughly indicates the median mass of the sample.}
    \label{figure:etameq}
\end{figure}

In \autoref{figure:etameq}, we show as black points the estimates of $\eta$ and $\meq$ with the respective uncertainties for the 8 clusters for which we could estimate both quantities. The coloured lines show \autoref{equation:BianchiniEta} for different values of $m$ from 0.3~\Msun to 1~\Msun as indicated by the legend. The line on which a cluster sits will be determined by both the mass range of the cluster sample, the frequency of masses within that range, and the relative PM uncertainties at different masses. With all this in mind, and given that the brightest stars tend to have the smallest uncertainties (and so pass the quality cuts with higher efficiency, despite being less numerous overall), we'd expect the clusters to sit at $\sim$ 0.7~\Msun with some scatter, and indeed we see that they lie between $\sim$-0.5--0.9~\Msun. Systems characterised by lower $\meq$ are closer to full equipartition and so have a higher $\eta$, and we see that our clusters follow this trend, as expected. Since these were two different fits to the same data, that the results agree with expectations lends confidence to the success of both fits.

But how do these values correlate with other cluster properties? We will consider correlations in both $\eta$ and $\meq$ as they are slightly different and one may prove more illuminating than the other.

In what follows, cluster ages are taken from \citet{VandenBerg2013}, with missing ages calculated as described in \autoref{section:masses}. All other cluster properties used are taken from \citet[][2010 edition]{Harris1996}. Where properties are provided or estimated in angular units, we convert them into physical units using the heliocentric distance from the same source.

Two-body relaxation, and the development of energy equipartition, depends on the number of interactions stars inside a cluster have experienced (see \autoref{section:introduction}). So an obvious correlation to look for is between $\eta$ or $\meq$ and the number of relaxation times the cluster has experienced, where $N = A / T$ for Age $A$ and relaxation time $T$. Relaxation time is not constant throughout a cluster; it decreases with increasing density, so tends to be shortest at the centre and to decrease with increasing radius. So we will consider both $\Ncore$ the number of core relaxation times and $\Nhalf$ the number of median relaxation times.

We may expect to find correlations with relaxation time directly as well, but with some scatter due to the different cluster ages (or similarly with age but with some scatter due to the different relaxation times). Similarly, we may expect to see correlations with the local velocity dispersion (a key ingredient in calculating the relaxation time), the core or half-light radii (which is related to the local density, which in turn correlates with relaxation time), the concentration (defined by the ratio of the core radius to the tidal radius), and even the ellipticity (a cluster will become rounder as it relaxes or as it experiences more relaxation times), but again with some scatter due to cluster-to-cluster differences in other properties. But, fundamentally, these correlations are all driven by $\Ncore$ and $\Nhalf$, so we elect to only show these two properties here.

\begin{figure*}
    \centering
    \includegraphics[width=\linewidth]{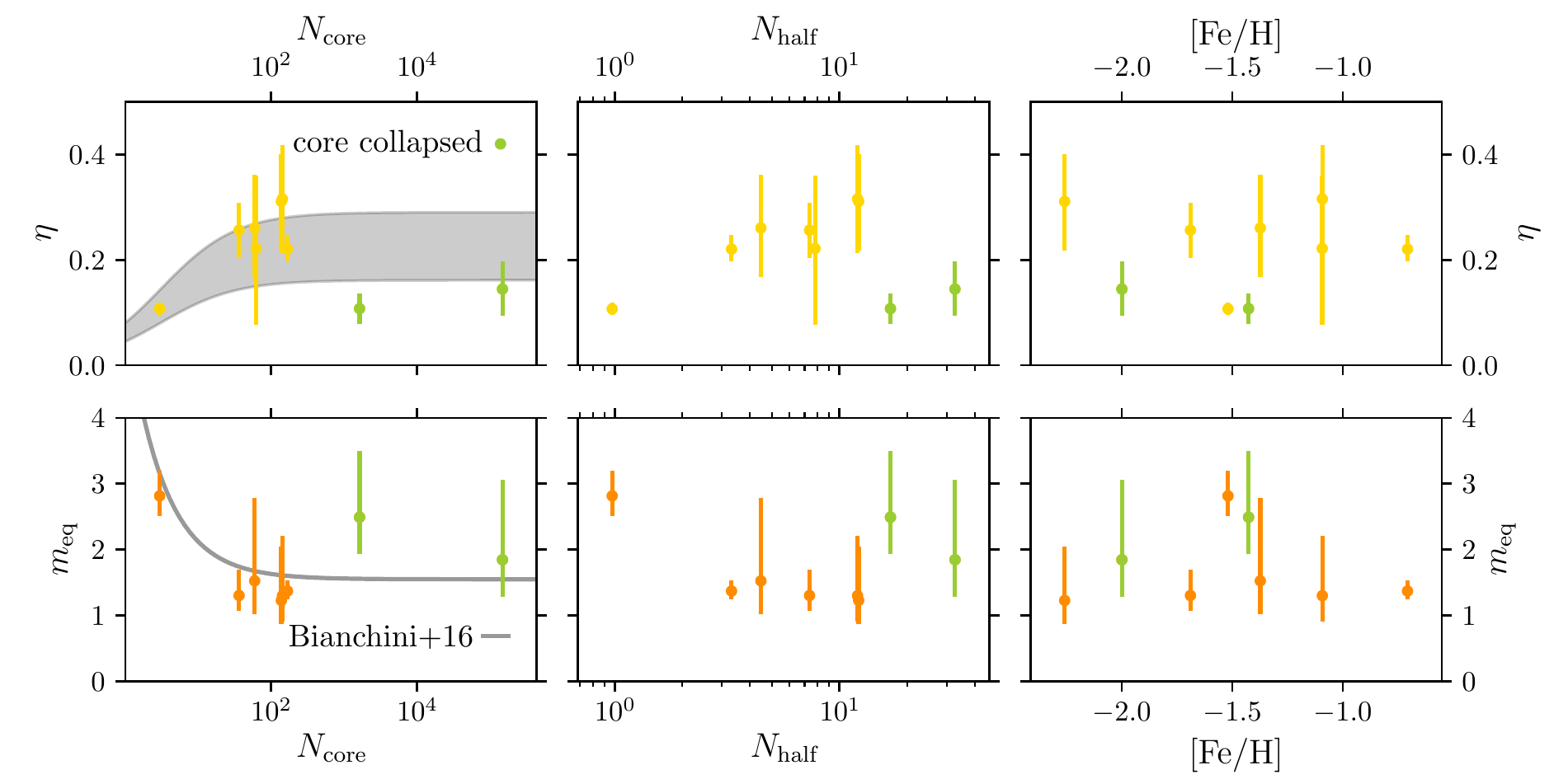}
    \caption{Correlations between equipartition parameter $\eta$ (top row) and equipartition mass $\meq$ (bottom row) with number of core relaxation times $\Ncore$ (left column), number of median relaxation times $\Nhalf$ (middle column), and metallicity [Fe/H] (right column). The yellow ($\eta$) and orange ($\meq$) points show the results for clusters not believed to be core collapsed. Green points highlight the two clusters believed to be core-collapsed (NGC\,6397 and NGC\,6752). The strongest correlations are observed with $\Ncore$, weaker correlations are also seen with $\Nhalf$. No correlation is observed with [Fe/H]. In the bottom-left panel, the solid line shows the trend line observed in simulated clusters by \citet{Bianchini2016} in non-core collapsed (orange) clusters. Our clusters agree with this trend line. The shaded region in the top-left panel shows this trend line for a range of different stellar masses. This panel is also provides some insight into how equipartition changes with time and the behaviour is consistent with that observed in simulated clusters by \citet{Trenti2013}.}
    \label{figure:etacorrelations}
\end{figure*}

\autoref{figure:etacorrelations} shows how $\eta$ (top row) and $\meq$ (bottom row) correlate with a number of different properties that we will discuss in turn. Clusters believed to be core collapsed (NGC\,6397 and NGC\,6752) are shown in green in all panels, but otherwise the panels showing $\eta$ have points coloured yellow and the panels showing $\meq$ have points coloured orange for consistency with previous plots.

The bottom-left panel shows the correlation between the equipartition mass $\meq$ and the number of core relaxation times $\Ncore$. In their study of simulated clusters, \citet{Bianchini2016} observed a very tight correlation between these two parameters and fitted a function to it, which we show here as the solid grey line. One key point is that they used only simulations that were not core collapsed, the equivalent to the orange points in this figure. Overall the orange points agree with the line very well and are thus consistent with the theoretical predictions, although given the uncertainties we cannot claim with statistical significance that this function fits the data or rule out the possibility of other trends.

But what about the core-collapsed clusters? \citet{Bianchini2018} used simulations of clusters that included core collapse to study how $\meq$ evolves both pre- and post-core collapse in a cluster. They observed that $\meq$ decreases steadily over time (or number of relaxation times, much as we see in the bottom-left panel of \autoref{figure:etacorrelations}) until core collapse, when $\meq$ experiences a sharp increase. This behaviour is certainly consistent with what we see here, albeit with only two clusters, with both of our core-collapsed (green) points sitting above the trend line.

We know that $\eta$ is correlated with $\meq$ from \autoref{equation:BianchiniEta} and \autoref{figure:etameq}. So we can invert the trend line from \citet{Bianchini2016} at constant mass to also predict how $\eta$ might change with $\Ncore$. These predictions are shown by the grey shaded region in the top-left panel of \autoref{figure:etacorrelations}. The upper limit of the region is the line for $m = 0.9$~\Msun and the lower limit is for $m = 0.5$~\Msun (with this choice of mass range motivated by \autoref{figure:etameq}). The region covers all masses in between. Again, we see that the non-core-collapsed clusters (in yellow) are in broad agreement with these predictions, while the core-collapsed clusters (in green) are a little offset. Just as we can only convert from $\meq$ to $\eta$ at constant mass (and so show a region instead of a single trend line here), the scatter of points in $\eta$ is larger than in $\meq$.

The second column of \autoref{figure:etacorrelations} shows $\eta$ and $\meq$ as a function of the median relaxation time $\Nhalf$. There is no theoretical trend line we can plot here\footnote{We attempted to fit a power law function, similar to that from \citet{Bianchini2016}, but unfortunately had too few points to successfully do so.}, but visually we can see that these panels have a broadly similar shape to the $\Ncore$ panels in the first column, which is not too surprising, including that the core-collapsed clusters are a little offset.

It is also apparent from the top-right figure why we saw two groups in the $\eta$ histograms in \autoref{figure:etahistogram} and $\meq$ histograms in \autoref{figure:meqhistogram}. There are 6 clusters with very similar $\Ncore$ values that have very similar $\eta$ or $\meq$ values. The other 3 clusters have lower $\eta$ values (higher $\meq$ values) either because they have experienced many fewer relaxation times and exist on a very different part on the fitted function, or because they are core collapsed and have experienced a sharp decrease in $\eta$ (increase in $\meq$ at that time. This creates two apparent groups due to the small number statistics but it is really a continuous underlying evolution.

Finally, the third column of \autoref{figure:etacorrelations} shows $\eta$ and $\meq$ as a function of cluster metallicity \FeH. In contrast to the other panels, here we see no correlation between either equipartition parameter and the metallicity.

\subsection{Core Collapse}
\label{section:corecollapse}

In the previous section, we stated that NGC\,6397 and NGC\,6752 are believed to be core collapsed; these judgements are based on their photometric properties. \citet{Bianchini2018} introduced a new statistic that identifies core-collapsed clusters based on their kinematic properties
\begin{equation}
    \ck = \frac{\meq \left( r < r_{50} \right)}{\meq \left( r_{50} \right)}
\end{equation}
where $r_{50}$ denotes the 50\% Lagrangian radius (that is the radius containing 50\% of the mass). The numerator is a `global' equipartition for all stars inside $r_{50}$ and the denominator is a `local' equipartition at $r_{50}$. $\ck > 1$ denotes core collapse. So how do our clusters fare under this prescription?

This is not straightforward to implement for a number of reasons. The first is that $r_{50}$ is defined by mass and in 3 dimensions. We only have projected radii on the plane of the sky, and do not have a full mass model with which to determine Lagrangian radii, only light profiles.

\citet{Libralato2018} estimated $\ck$ in their study of NGC\,362, using the projected radii and half-light radius $\Rhalf$ in place of $r_{50}$, so we could consider the same. This brings us to the second problem we face with our sample: some clusters do not have data all the way out to $\Rhalf$. \citet{Bianchini2018} showed that their statistic also worked when considering $r_{40}$ or $r_{60}$ in place of $r_{50}$ -- explaining why \citet{Libralato2018} were able to safely use $\Rhalf$ in place of $r_{50}$ even though mass does not follow light in GCs. This suggests there is some flexibility in the radius at which the statistic is evaluated and that data out to a significant fraction of $\Rhalf$ may be sufficient. This mitigates the issue for some of our clusters, but some so far have data only inside their $\Rhalf$ it's not clear that the statistic could be measured reliability.

There is a third problem. Calculating $\meq$ ideally requires that the mass range be constant over the range of radii covered. We know that dispersion changes as a function of both mass and radius so, for example, having only high-mass stars at small radii and low-mass stars at large radii would tend to wash out any equipartition signal. When calculating a local $\meq$ the radial range is small, so the mass range is constant and this is not an issue. But, when calculating a global $\meq$ this can be a serious problem. Indeed we tend to find that the mass range in the intermediate regions covered by our catalogues is that of the whole dataset, but that in the inner and outer regions of the catalogues there is significant incompleteness as a function of mass, sometimes at the high-mass end, sometimes at the low-mass end, and sometimes at both. Depending on the details of the incompleteness this can either inflate or wash out any signal of equipartition, and certainly lead to incorrect values of $\meq$ in the $c_k$ equation.

This problem is the most critical, as it is very hard to account for. The completeness as a function of mass is a complicated function of the magnitude of the stars, the density of the clusters, and the details of the observations that were used to measure the proper motions.

Although we tried to estimate $c_k$ for our clusters, in the end these 3 problems meant that we were unsuccessful. This remains an intriguing statistic, and being able to estimate core collapse from kinematics would be both valuable and a more faithful representation of the underlying physics, but is beyond the limits of the data at this time.

\subsection{Multiple Populations}

Recent theoretical studies suggest that first- and second-generation (1G and 2G) stars are at different stages of energy equipartition \citep[e.g.][]{Vesperini2021}. In dynamically young clusters (those with long relaxation times), these differences could still be preserved. This is also borne out by observations; \citet{Bellini2018} found that the 1G and 2G stars in NGC\,5139 have different degrees of equipartition (with neither being even close to full equipartition). If the 1G and 2G stars have different spatial concentrations as well as different kinematics, then this could result in different (global) average $\eta$ values for the inner and outer parts of a cluster. It is also worth noting that \citet{Libralato2019} found no energy equipartition differences between 1G and 2G stars in NGC\,6352, but attribute this to its advanced evolutionary state having washed out any differences.

While interesting to explore, separating the samples into multiple populations adds an additional layer of complexity that goes beyond the scope of this paper. Here our driver was to understand the global equipartition properties, and we've already shown that this analysis pushes the limits of what is possible with the present catalogues.

\subsection{Previous Studies}

We have already touched on how our results compare with some previous studies, but we will do so in more detail here. Our estimates and other observational studies have calculated projected values. The simulations we reference here also made predictions in projection so may be directly compared to our observations.

Let us begin with the theoretical predictions of \citet{Trenti2013}, which were that 1) clusters do not reach full equipartition, 2) the maximum $\eta$ reached is $\sim 0.2$, and 3) the central regions of the cluster show a strong initial increase in $\eta$ but then turn over and decrease to $\eta \sim 0.1$.

Our results support the first prediction: none of the clusters studied here has reached full equipartition. Three clusters support the second prediction, with estimates for $\eta < 0.2$, while the other 6 clusters have values higher than predicted in the range $0.2 < \eta < 0.35$. The third prediction is harder for us to study directly as we have a single snapshot of multiple clusters and not time evolution of a single cluster. However, let us consider the upper-left plot of \autoref{figure:etacorrelations}, which shows $\eta$ as a function of the number of relaxation times $\Ncore$ experienced by the cluster. Assuming that all clusters follow similar trends, as \citet{Trenti2013} showed, then we can use this plot as a proxy for the time evolution of $\eta$. And indeed, we see that $\eta$ starts with a sharp rise, hits a maximum and then decreases again to $\eta \sim 0.1$, in good agreement with the predictions.

We can also compare results for some individual clusters. As discussed in \autoref{section:introduction}, \citet{Anderson2010} estimated $\eta \sim 0.2$ in NGC\,5139 ($\omega$ Centauri) using HST PMs measured over a four-year baseline. \citet{Trenti2013} quoted a revised estimate of $\eta \sim 0.16$ using an earlier version of the catalogue that we used in this study. Here, we found $\eta = 0.107 ^{+0.013} _{-0.012}$, in reasonable agreement, although a little lower, than previous results.

\citet{Heyl2017} studied mass segregation and energy equipartition in NGC\,104 using \textit{HST} PMs. In their outer field, $\meq \approx 2$~\Msun. In their inner field, they find $\meq \approx 1$~\Msun. In a region intermediate to the ones they spanned, we find $\meq = 1.37^{+0.16}_{-0.13}$~\Msun, also intermediate to their values, which is broadly consistent with their results.

Unfortunately we are unable to make a direct comparison to the NGC\,362 study of \citet{Libralato2018}. Although NGC\,362 was one of the clusters in our initial sample of 22 clusters, it was rejected at the local dispersion cleaning stage. However, we can consider how that study fits in with our broader results. \citet{Libralato2018} found a global $\eta = 0.114 \pm 0.012$, within the range we estimated for the 9 clusters in our study. This value would put NGC\,362 in the lower $\eta \sim 0.1$ peak that we observed in \autoref{figure:etahistogram}.

They also determined an $\eta$ profile with radius for NGC\,362 that decreases from $\eta \sim 0.4$ near the cluster centre to $\eta \sim 0.1$ towards the edge of the field. We were unable to draw any strong conclusions from our $\eta$ profiles, although they tentatively suggested that the clusters either have a flat $\eta$ profile or even an $\eta$ that slightly increases with radius (over the range of the data). The quality of the NGC\,362 PMs is clearly superior to those in this study, so this tentative inconsistency likely speaks to our earlier conclusion that calculating $\eta$ profiles is at the very limit of what we can do with the present catalogues.

While this paper was being finalised, \citet{Pavlik2021} published a study of the interplay between velocity anisotropy and energy equipartition, again in \textit{N}-body simulations of globular clusters. Before core-collapse, they found that the degree of equipartition exhibited by the radial and tangential velocities differs in the intermediate and outer regions of their clusters, with the effect becoming more significant at large radii. In fact, they showed that equipartition can even be inverted in the outer parts of the clusters. They also showed that the evolution towards energy equipartition depends on the degree of initial anisotropy.

Studying equipartition separately in radial and tangential PMs and considering $\eta$ and anisotropy together is beyond the scope of this paper. Although certainly interesting to consider in future work, arguably these catalogues are not best suited to directly address the conclusions in \citet{Pavlik2021}. Firstly, we noted in \citetalias{Watkins2015a} that the clusters are isotropic or show only mild radial anisotropy over the range spanned by the data. And secondly, the effects \citet{Pavlik2021} observed were most noticeable or only present in the intermediate and outer regions of their simulated clusters, whereas our catalogues are restricted to the central regions.

\subsection{Future Prospects with \textit{HST} and \textit{Gaia}}

In the previous section, we noted that we are unable to make a direct comparison to the NGC\,362 study of \citet{Libralato2018}. The latter study was actually based on an improved version of the catalogue that we used here. We elected to use the \citet{Bellini2014} version of the catalogue both because we wanted to perform a uniform analysis on a uniform set of catalogues, and because the \citet{Libralato2018} version of the catalogue had already been studied for equipartition in that paper.

The good news is that NGC\,362 was the first in a series of improved catalogues that will be forthcoming. The improved catalogues have additional steps in the PM determination that make them more accurate and decrease their uncertainties \citep[see][for details]{Libralato2018}. In the intervening years, more data has become available for many clusters in the original \citet{Bellini2014} sample; more epochs and longer baselines will further improve the accuracy of the catalogues for these clusters. Finally, there are clusters for which no PM determination was possible when the original catalogues were compiled, but which can now be included.

The difference between the (failed) NGC\,362 study in this work and the (successful) study in \citet{Libralato2018} highlights the potential for the improved catalogues to both build on the present study and provide new insights into equipartition in GCs.

One limitation of these catalogues is their spatial coverage. We can  measure PMs only in the closest clusters for which the photometric accuracy is high, and the angular motions are measurable over the typical 3-10 year baselines that exist for much \textit{HST} data. These clusters extend far beyond the typical \textit{HST} field of view. So \textit{HST} imaging (of any depth or quality) typically only exists for GC centres, or at most in the centre and for a single off-centre field in any given GC.

To study equipartition in GCs beyond these central fields, we will need to turn to \textit{Gaia}. This will be challenging. With a limiting magnitude of $G \sim 21$~mag, \textit{Gaia} PMs exist only for stars above the MSTO in most clusters, severely restricting the range of stellar mass spanned. Blue stragglers will offer the best chance to study equipartition in such cases. For very close clusters, \textit{Gaia} does go deep enough to cover the upper end of the main sequence and provide a feasible range of stellar masses. At present, the \textit{Gaia} PM accuracy is such that kinematic studies are possible but challenging even for the brighter stars. Stars fainter than $G \sim 19.5$~mag typically have PM uncertainties too large to be usable. However, with more epochs, longer baselines, and proper treatment of crowded fields \citep[][a particular concern in dense GCs]{Pancino2017}, PM accuracy will continue to improve. As \citet{Pancino2017} beautifully demonstrated, only later \textit{Gaia} data releases that include a sufficient number of well-measured stars at or below the MSTO are expected to be transformative for GC research.

Together, \textit{HST} and \textit{Gaia} are highly complementary. At present, \textit{HST} offers a unique opportunity to study kinematics along the main sequence. In the future, \textit{HST} will continue to provide the best PM data to study the central regions (even with treatment of crowded regions applied to the \textit{Gaia} data) and will provide PMs for stars below \textit{Gaia}'s magnitude limit, but \textit{Gaia} will be able to study the outer regions that \textit{HST} does not cover.

The legacy of PM studies with \textit{HST} will likely be continued with the \textit{James Webb Space Telescope} (\textit{JWST}) and the \textit{Nancy Grace Roman Space Telescope} (\textit{Roman}). \textit{JWST} will go deeper than \textit{HST}, enabling us to push equipartition studies to more distant clusters and to increase the mass range spanned in some of the existing clusters, but with a field of view comparable to \textit{HST}, it will not expand our knowledge of the outer parts of clusters. By contrast, \textit{Roman} will be able to cover most GCs out to their tidal radii and beyond in a single pointing, so will greatly expand our studies of their outer regions \citep{Bellini2019}. But it will take time for both observatories to build up significant baselines.

Even with future improvements in PM quality and coverage, PMs cover only 2 dimensions of motion. Ideally, we would study kinematics in all 3 dimensions to fully characterise a GC. At present, the best GC LOSVs come from the Very Large Telescope / Multi Unit Spectroscopic Explorer \citep[VLT/MUSE, see][]{Kamann2018}, but equipartition studies remain out of reach even with these data sets. However, forthcoming instruments, such as the Extremely Large Telescope's High Angular Resolution Monolithic Optical and Near-infrared Integral field spectrograph\footnote{\url{https://elt.eso.org/instrument/HARMONI/}} \cite[ELT/HARMONI,][]{Thatte2021}, offer hope for the future.

\section{Conclusions}
\label{section:conclusions}

We have studied the stellar kinematics as a function of stellar mass in 9 Galactic GCs using proper motions measured with \textit{HST}, and in so doing, studied their degree of energy equipartition.

We began with 22 catalogues. For these, we assigned stellar masses to each star via isochrone fitting and then performed a series of cuts designed to remove poor quality stars. 6 catalogues failed the quality selection cuts and were removed at this stage. We then performed further cuts designed to remove ``noisy" stars and improve the effective signal-to-noise of the sample. A further 7 clusters were removed at this stage.

This left us with 9 cleaned, high-quality catalogues to study. To these, we fit two functions to the velocity dispersion as a function of stellar mass: the first a simple power law in which degree of equipartition was parameterised by power-law index $\eta$, and the second from \citet{Bianchini2016} in which the degree of equipartition was parameterised by equipartition mass $\meq$. We obtained good fits for all 9 clusters using the power-law function, and for 8 of the 9 clusters using the \cite{Bianchini2016} function.

Using the 8 clusters for which both fits were successful, we performed statistical tests to determine which was the better-fitting function and found that 4 preferred the power-law and four preferred the \citet{Bianchini2016} fit, all at very low significance. With the present data it is not possible to determine which function is best.

The power-law fits produced $\eta \sim 0.1-0.35$. This is consistent with previous theoretical work by \citet{Trenti2013} who predicted that clusters never reach full equipartition. Although our maximum $\eta$ observed is higher than their maximum prediction of $\eta \sim 0.25$.

We looked for correlations between $\eta$ or $\meq$ and various cluster parameters. The strongest correlations were between $\eta$ or $\meq$ and $\Ncore$ or $\Nhalf$, the number of core or median relaxation times. \citet{Bianchini2016} found a very strong theoretical correlation between $\meq$ and $\Ncore$ in their theoretical work to which they fitted a line. Our data were in very good agreement with their trend line. The correlations between $\eta$ and $\Ncore$ are also consistent with the predictions of \citet{Trenti2013}. We noted correlations with other cluster structural and kinematic properties, but all drive or are driven by the number of relaxation times. We found no correlation with metallicity.

We also investigated how $\eta$ changes as a function of distance from the cluster centre; most clusters were consistent with being flat, and two shows mild hints of increasing with radius, but the results were of low statistical significance, indicating that this analysis is at the limits of what is possible with the present data.

\begin{acknowledgments}

We thank the referee for a helpful report that improved the presentation of this work.

LLW wishes to thank Roger Cohen, Erik Tollerud, and Elena Sacchi for very useful conversations.

Support for this work was provided by grants for \textit{HST} program AR-15055 provided by the Space Telescope Science Institute, which is operated by AURA, Inc., under NASA contract NAS 5-26555.

This project has received funding from the European Research Council (ERC) under the European Union's Horizon 2020 research and innovation programme under grant agreement No 724857 (Consolidator Grant ArcheoDyn).

This research made use of Astropy\footnote{\url{http://www.astropy.org}}, a community-developed core Python package for Astronomy. This research has made use of NASA's Astrophysics Data System Bibliographic Services.

This project is part of the HSTPROMO (High-resolution Space Telescope PROper MOtion) Collaboration\footnote{\url{http://www.stsci.edu/~marel/hstpromo.html}}, a set of projects aimed at improving our dynamical understanding of stars, clusters and galaxies in the nearby Universe through measurement and interpretation of proper motions from \textit{HST}, \textit{Gaia}, and other space observatories. We thank the collaboration members for the sharing of their ideas and software.
\end{acknowledgments}

\vspace{5mm}

\facility{HST}

\software{
	Astropy \citep{astropy2013, astropy2018},
	\textsc{emcee} \citep{ForemanMackey2013},
	IPython \citep{ipython2007},
	Matplotlib \citep{matplotlib2007},
	NumPy \citep{numpy2020},
	SciPy \citep{scipy2020}
}

\bibliography{refs}

\appendix

\section{Data Tables}
\label{appendix:datatable}

\begin{deluxetable}{ccccc}
\tablecaption{Binned velocity dispersion profiles as a function of stellar mass. \label{table:dispersions}}
\tablehead{
  \colhead{Cluster} &
  \colhead{Bin} &
  \colhead{$N_\star$} &
  \colhead{$m$} &
  \colhead{$\sigma$} \\
  \colhead{} &
  \colhead{} &
  \colhead{} &
  \colhead{(\Msun)} &
  \colhead{(mas/yr)}
}
\startdata
NGC\,104  & 1 & 1642 & 0.4021 $\pm$ 0.0635 & 0.6256 $\pm$ 0.0166 \\
          & 2 & 1640 & 0.5118 $\pm$ 0.0172 & 0.6092 $\pm$ 0.0107 \\
          & 3 & 1640 & 0.5597 $\pm$ 0.0115 & 0.5931 $\pm$ 0.0111 \\
          & 4 & 1642 & 0.6013 $\pm$ 0.0123 & 0.5669 $\pm$ 0.0102 \\
\enddata
\tablecomments{Columns: (1) Cluster, (2) bin number, (3) number of stars in bin, (4) average mass of stars in bin, (5) velocity dispersion in bin.}
\end{deluxetable}

Here we provide the binned velocity dispersion profiles as a function of stellar mass, plotted as the black points with errorbars in \autoref{figure:1dClassicFits} and \autoref{figure:1dBianchiniFits}. A portion is provided here to show form and content, the full version will be available through the journal, or can be obtained by request before publication.

\end{document}